\documentclass{aa}
\usepackage{astron,times,txfonts,latexsym,graphics,graphicx}

\begin{document}

\title{The Calibration of Str\"omgren $uvby$-H$\beta$ 
       Photometry for Late-Type Stars 
       -- a Model Atmosphere Approach}
\author{Anna \"Onehag\inst{1} 
\and    Bengt Gustafsson\inst{1}
\and    Kjell Eriksson\inst{1}
\and    Bengt Edvardsson\inst{1} }

\institute{Department of Physics and Astronomy, 
           Uppsala Astonomical Observatory,
           Box 515, S-751\,20 Uppsala, Sweden}

\date{Received / Accepted}
\offprints{Anna \"Onehag,
\email{Anna.Onehag@fysast.uu.se}}

\authorrunning{\"Onehag et al.}
\titlerunning{Calibration of Str\"omgren Photomtery}

\abstract
{ 
 The use of model atmospheres for deriving stellar fundamental parameters, 
 such as $T_{\rm eff}$, $\log g$ and [Fe/H], will increase as we find and explore 
 extreme stellar populations where empirical calibrations are not yet available. 
 Moreover calibrations for upcoming large satellite missions of new
 spectrophotometric indices, similar to the $uvby-$H$\beta$ system, will be needed.
}
{We aim to test the power of theoretical calibrations based on a new generation 
 of MARCS models by comparisons with observational photomteric data. 
 }
{We calculate synthetic $uvby-$H$\beta$ colour indices from synthetic spectra.
 A sample of 388 field stars as well as stars in globular clusters is used for 
 a direct comparison of the
 synthetic indices versus empirical data and for scrutinizing the possibilities of
 theoretical calibrations for temperature, metallicity and gravity.
 }
{We show that the temperature sensitivity of the synthetic $(b-y)$ colour
 is very close to its empirical counterpart, whereas the temperature scale based
 upon H$\beta$ shows a slight offset. The theoretical
 metallicity sensitivity of the $m_{1}$ index (and for G-type stars its combination with $c_1$)
 is somewhat larger than the empirical one, based upon spectroscopic determinations.
 The gravity sensitivity of the
 synthetic $c_1$ index shows a satisfactory behaviour when compared to obervations of F stars.
 For stars cooler than the sun a deviation is significant in the $c_{1}$--$(b-y)$ diagram.
 The theoretical calibrations of $(b-y)$, $(v-y)$ and $c_{1}$ seem to work well for Pop II
 stars and lead to effective temperatures for globular cluster stars supporting recent
 claims by Korn et al. (2007) that atomic diffusion occurs in stars near the turnoff 
 point of NGC 6397.
 }
{Synthetic colours of stellar atmospheres can indeed be used, in many cases,
 to derive reliable fundamental stellar parameters. The deviations seen
 when compared to observational data could be due to
 incomplete linelists but are possibly also due to effects of
 assuming plane-parallell or spherical geometry and LTE.
 }


\maketitle

\section{Introduction}
   The $uvby$ photometric intermediate-band system of Str\"omgren (1963) and the H$\beta$ 
   narrow-band system (Crawford 1958, 1966) were combined early and soon proved to be
   a most powerful means of determining fundamental parameters of stars.
   Calibrations of the systems were also developed early by Str\"omgren, Crawford
   and collaborators. Generally, semi-empirical methods were used with the calibrations
   made by means of sets of stars with parameters determined in more fundamental ways.
   The first metallicity calibration 
   of the $m_1$ index for solar-type stars by Str\"omgren (1964)
   was thus based on spectroscopical [Fe/H] values of Wallerstein (1962) and the first
   luminosity calibration of the $c_1$ index by Str\"omgren (see Crawford 1966) used
   cluster stars in the main-sequence band. This general semiempirical approach continued 
   with calibrations like those of the $m_1$ index by Nissen (1970), Gustafsson $\&$ Nissen 
   (1972),
   Nissen $\&$ Gustafsson (1978) and Nissen (1981) where very-narrow-band spectrophotometry of
   groups of weak lines was calibrated with synthetic spectra. These stars were then used as
   calibration stars for 
   $uvby$-H$\beta$ photometry. Other calibrations follow this semiempirical 
   approach e.g. those of Crawford (1975), Ardeberg $\&$ Lindgren (1981) and Olsen (1988) 
   for G and K-type stars, as well as by Schuster \& Nissen (1989) for metal-poor stars.  
   In the calibration by Alonso et al. (1996) the effective temperature 
   calibration was based on stellar temperatures from the Infrared Flux Method. The 
   calibrations by Nordstr\"om et al. (2004) and Holmberg et al. (2007)  
   additionally utilised Hipparcos parallaxes for the surface gravity calibration.

   Another development towards calibration of the systems also started early: the direct 
   calculation of
   photometric indices by means of model atmospheres. Such theoretical calibrations were
   attempted for early-type stars with relatively line-free spectra. For late
   type stars, a statistical correction for the effects of spectral lines was made by Baschek
   (1960) in his calibration of the Str\"omgren $m$ index (a predecessor to $m_1$).  A first 
   systematic and detailed
   calculation of $uvby-$H$\beta$ indices for a grid of F and G dwarf model atmospheres was 
   published by Bell (1970), using scaled solar model atmospheres. Bell $\&$ Parsons (1974) 
   calculated $uvby$ colours for flux-constant model atmospheres of F and G supergiants, 
   while Gustafsson $\&$ Bell (1979) produced theoretical colours in a number of systems, 
   including the $uvby$ system, for a grid of giant-star model atmospheres.
   Relya \& Kurucz (1978) calculated $uvby$ and $UBV$ colours from early ATLAS models, and 
   discussed their shortcomings for late-type stars. $uvby$ colours for new
   sets of Kurucz models were published by Lester, Gray $\&$ Kurucz (1986).
   Castelli $\&$ Kurucz (2006) published H$\beta$ indices. Sometimes, semiempirically 
   corrected fluxes from model 
   atmospheres have also been used for calibrations of Str\"omgren photometry, see e.g. 
   Lejeune et al. (1999) and Clem et al. (2004).   

   The need for reliable calibrations of $uvby-$H$\beta$ photometry has increased in the last 
   decade, not the least for estimating parameters of new and more "exotic" stars, such 
   as very metal-poor and super-metal-rich stars which are not found at great abundance in the 
   solar neigbourhood so that relatively complete sets of calibration stars cannot 
   easily be established. Furthermore, the preparation for the Gaia satellite includes a 
   careful analysis of the power of model atmospheres to provide a detailed astrophysical 
   calibration of the photometric system of the satellite. This analysis needs support 
   by a detailed test of the problems and possibilities to make a detailed theoretical 
   calibration of, e.g., the $uvby-$H$\beta$ photometry. 
   
   Subsequently, we shall present the theoretical models and colours (Sect. \ref{sec:models} 
   and \ref{sec:colours}). Stellar samples for empirical comparisons are discussed in 
   Section \ref{sec:stellsamp}. Next, the
   discussion will be focused on the determination of effective temperature, metallicity and 
   surface gravity of the stars, by discussing the 
   calibration of $(b-y)$ and H$\beta$ indices (Sect. \ref{sec:teffcalib}, effective temperature), 
   of $m_1$ (Sect. \ref{sec:metcalib}, metallicity) and $c_1$ 
   (Sect. \ref{sec:gravcalib}, gravity) indices, devoting more limited interest to 
   "secondary" effects such as the
   metallicity sensitivity of $(b-y)$ and $c_1$, or the gravity sensitivity of $m_1$.
   In each section the results will be compared with empirical and 
   semi-empirical data and calibrations. Finally, in the last Section some comments will be made on
   the success and the problems of the theoretical calibrations, conclusions will
   be drawn and recommendations given. 

\section{Model atmospheres and calculated spectra}
\label{sec:models}
   The theoretical tools used in modelling the stellar colours are model stellar
   atmospheres and their calculated fluxes. These are based on extensive
   atomic and molecular data. Here, we shall briefly present the models and 
   data used and refer to more complete descriptions. 
   
\subsection{Model atmospheres}
    The stellar atmosphere code MARCS (Gustafsson et al. 2008,
    {\it http://marcs.astro.uu.se}) was used to 
    construct a grid of 168 theoretical 1D, flux constant, radiative + mixing-length 
    convection, LTE models with fundamental atmospheric parameters as follows: 
    $T_{\rm eff}$ = 4500, 5000, 5500, 6000, 6500 \& 7000\,K,
    $\log g$ = 2.0, 3.0, 4.0, 4.5 and 
    [Me/H] = 0.5, 0.0, --0.50, --1.0, --2.0, --3.0, --5.0, [Me/H] denoting the logarithmic
    over-all metallicity with respect to the sun.
    Plane parallell (ppl) stratification was assumed for $\log g$\,=\,4.5, 
    4.0 \& 3.0, whereas spherical (sph) symmetry was assumed for $\log g$\,=\,2.0.
    For the spherically symmetric models a mass of 1\,M$_{\odot}$ was adopted. The local
    mixing-length recipe was used to describe convective fluxes, for more details see 
    Gustafsson et al. (2008).
    
    Elemental abundances were adopted from Grevesse and Sauval (1998) except for the CNO 
    abundances which were adopted following Asplund et al. (2005).

 \subsection{Synthetic spectrum calculations} 
  \label{sec:synthspeccalc}
    In order to calculate synthetic spectra of sufficiently high resolution the 
    Uppsala BSYN code was used with the MARCS models as input. The spectra were 
    calculated within the wavlength limits of the Str\"omgren $uvby$ filters with 
    wavelength steps of 0.02\,\AA.
    The mictroturbulence parameter, $\xi_{t}$, was set to 1.0\,km\,s$^{-1}$ and 
    1.7\,km\,s$^{-1}$, when calculating spectra based on ppl models and sph models, 
    respectively. The effects of changing $\xi_{t}$ are 
    studied in Section \ref{sec:metcalib}.

 \subsection{Line lists}
    We collected atomic line data from the Vienna Atomic Line Data Base, VALD (version I,  
    Kupka et al. 1999). 
    For hydrogen line data a version of the code HLINOP was used, and has been described by
    Barklem \& Piskunov (2003). This code has been developed  based on the
    original HLINOP by Peterson \& Kurucz (see http://
    kurucz.harvard.edu/).  The hydrogen line profiles are calculated
    including Stark broadening, self-broadening, fine structure, radiative
    broadening, and Doppler broadening (both thermal and turbulent).  The
    Stark broadening is calculated using the theory of Griem (1960 and
    subsequent papers) with corrections based on Vidal et al. (1973).
    Self-broadening is included following Barklem et al (2000) for H$\alpha
    $, H$\beta$ and H$\gamma$, while for other lines the resonance
    broadening theory of Ali \& Griem (1966) is used.
    For molecules, data for C$_{2}$ including $^{12}$C$^{13}$C lines were gathered 
    from Querci, Querci \& Kunde (1971) and Querci (1998, private communication with B. Plez), 
    except for the Fox-Herzberg band in the UV for which data was taken from Kurucz (1995).
    Molecular data for CH with $^{13}$CH comes from Plez et al. (2008) and Plez (2007, priv. 
    communication).
    Data for the CN molecule with $^{12}$C$^{15}$N, $^{13}$C$^{14}$N, and $^{13}$C$^{15}$N  
    are also from Plez (priv. communication).
    CO data with $^{13}$C$^{16}$O was gathered from Kurucz (1995) 
    as well as NH with $^{15}$NH, OH with $^{18}$OH, and MgH with $^{25}$MgH and $^{26}$MgH.  
    TiO was not taken into account in the calculations since absorption by this molecule is 
    expected to have small effects on the spectra within the effective-temperature range 
    considered.

\section{Colour index calculations}
\label{sec:colours}

\subsection{Filter profiles}
\begin{figure}[h!]
     \resizebox{\hsize}{!}{\includegraphics[angle=90]{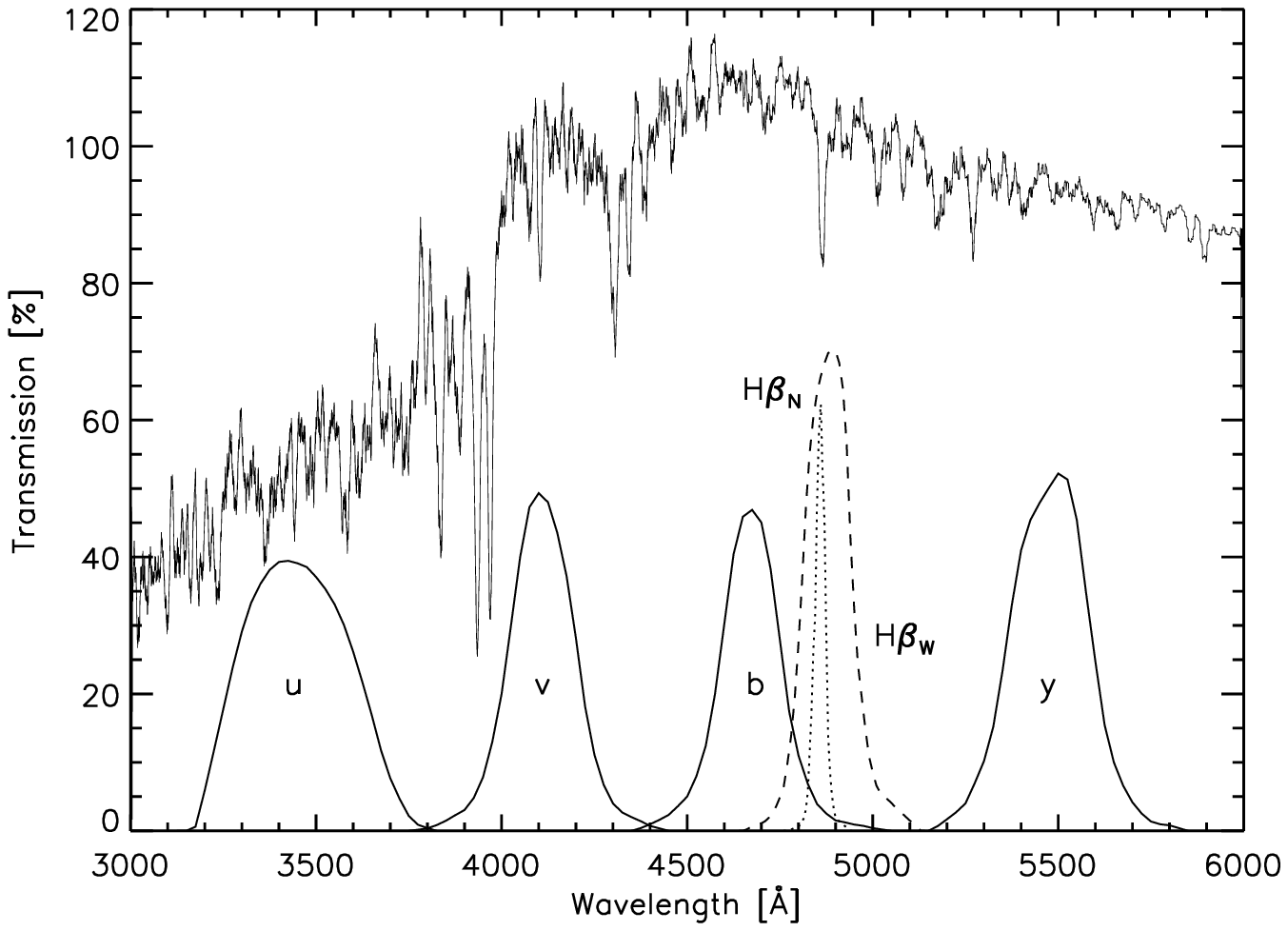}}
     \caption{The $uvby-$H$\beta$ transmission functions of the standard systems plotted 
       as a function of wavelength.
       As a comparison, the flux (per \AA ngstr\"om  unit) of a model with 
       $T_{\rm eff}$\,=\,6000\,K, $\log g$\,=\,4.0 and [Me/H]\,=\,0.0 is plotted on an 
       arbitrary flux scale.}
     \label{fig:transm}
    \end{figure}

    To determine the theoretical $m_{1}$, $c_{1}$ and $(b-y)$ indices, transmission profiles 
    of the Str\"omgren $uvby$ filters (Crawford \& Barnes (1970) see Fig. \ref{fig:transm}) were 
    multiplied with the calculated model stellar surface flux within the wavelength range 
    of the filters:
\begin{eqnarray}
    mag\,=\,-2.5\,{\rm \log}\left(\frac{\int{F_{\lambda}T_{\lambda}d\lambda}}
                                               {\int{T_{\lambda}d\lambda}}\right)+const. \nonumber
\end{eqnarray}
    where $F_{\lambda}$ and $T_{\lambda}$ are the flux and the relevant transmission profile,
    respectively.
    The theoretical magnitudes were converted into colour indices via the definitions:         
    $c_{1}\equiv u - 2v + b,\,\,m_{1}\equiv v - 2b + y. $
                      
    The H$\beta$ index is defined (Crawford 1958) as the ratio of the flux measured through 
    a narrow and a wide profile, respectively, both centered around the H$\beta$ line:
\begin{eqnarray}
    {\rm H}\beta\,=\,-2.5\left(\log\frac{\int{F_{\lambda}T_{N,\lambda}d\lambda}}
                                        {\int{T_{N,\lambda}d\lambda}}
                               -         
                               \log\frac{\int{F_{\lambda}T_{W,\lambda}d\lambda}}
                                        {\int{T_{W,\lambda}d\lambda}}\right) + const.
\nonumber
\end{eqnarray}   
    where $T_{W,\lambda}$ and $T_{N,\lambda}$ are the wide and the narrow filter profiles, 
    respectively. We have found notable differences between
    synthetic H$\beta$ indices calculated with the Kitt Peak filter set denoted (9,\,10) and those 
    calculated with the (212,\,214) set, both described by Crawford \& Mander (1966, CM66).
    Here, the Kitt Peak (212,\,214) filter system was 
    chosen (Fig. 1 CM66) with a transmission function adopted from Castelli \& 
    Kurucz (2006), see Figure \ref{fig:transm}.
    
  \subsection{Transformation to the observational system}
     The frequent use of filter-profiles different from those that originally have defined
     the photomteric system needs some extra considerations. For the H$\beta$ index this mainly 
     comes in at the zero-point determination. Since the observed indices are defined 
     on a system with zero-points set by particular standard stars, we must apply 
     corresponding zero-point shifts to our theoretical system. The much observed, and 
     frequently used, star Vega was chosen to estimate these zero-points.
     Thus, a model atmosphere for Vega was calculated with parameters $T_{\rm eff}$ = 9550\,K,
     $\log g$ = 3.95 and [Me/H] = --0.5 (mean values of selected measurements presented
     in SIMBAD), and a synthetic spectrum was computed.
     Theoretical indices for the Vega model were calculated and compared to 
     observed values: $c_{1}$\,=\,1.088, $m_{1}$\,=\,0.157 and $(b-y)$\,=\,0.003 (Hauck and 
     Mermilliod 1998). The resulting differences between the theoretical and 
     observed colours were then added as constants to all calculated Str\"omgren colours in 
     our model grid.\\
     For the H$\beta$ index, Crawford \& Mander (1966) presented several filter systems, 
     of which we choose the (212,\,214) filters as mentioned above.  
     Indices calculated by using this filter set, which are referred to as H$''\beta$ below, 
     should be transformed via  a set of equations (CM66), in order to agree 
     with previous Crawford--Mander observations on their
     standard system. The transformation for the (212,214) filters
     is described by the following two equations (CM66, Table\,III):\\\\
     H$\beta = 0.374 + 1.305\,$H$'\beta$,\,\, B stars\\\\
     H$\beta = 0.248 + 1.368\,$H$'\beta$,\,\, A,\,F stars,\\\\
     derived from a set of 45 and 35 bright stars, where H$'\beta$ and H$\beta$
     are the observed and tranformed indices, respectively. 
     The later equation for A and F stars was applied on all H$\beta$
     indices in our theoretical grid. 
     
\begin{figure}[h!]
       \resizebox{\hsize}{!}{\includegraphics[angle=90]{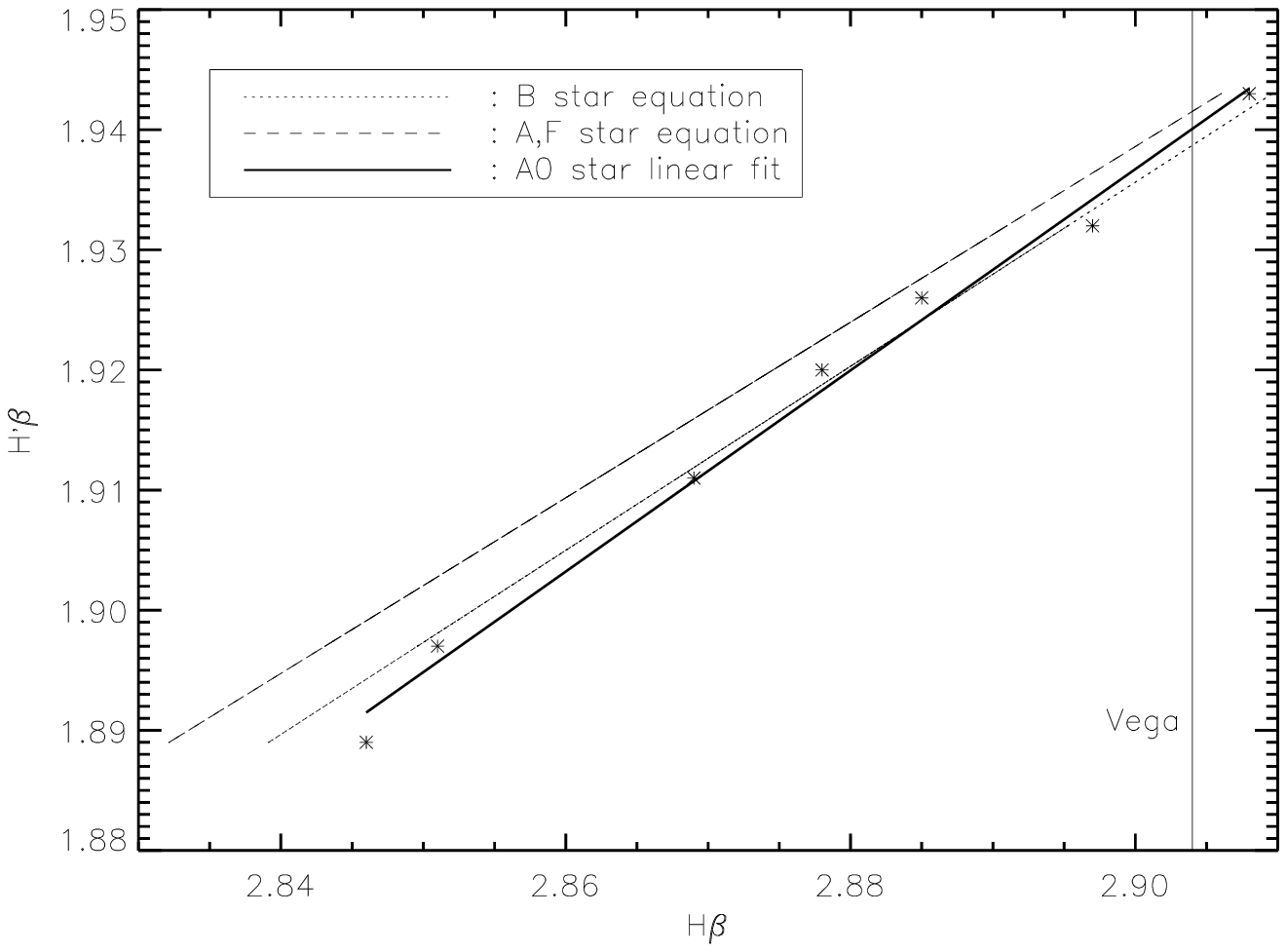}}
       \caption{H$\beta$ and H$'\beta$ values for A0 stars tabulated in CM66 are plotted individually 
                and represented by a linear regression (thick line), as well as the 
                transformation relations for B and A+F stars, respectively, following CM66 (thin lines).}
      \label{fig:betatransf}
     \end{figure}

     The A0 star Vega clearly poses a number of problems for determining the zero-point of 
     the $uvby$ and H$\beta$ indices. It is known to be rapidly rotating, but with its axis close
     to the line of sight (Gulliver, Hill \& Adelman 1994, Hill, Gulliver \& Adelman 2004). 
     Vega has also been regarded to show mild $\lambda$ Bootis star characteristics,
     such as certain non-solar abundance ratios as well as dust emission in the IR
     (see Gigas 1988, Hill 1995, Ilijic et al. 1998, Adelman \& Gullliver 1990, Heiter, Weiss \& 
     Paunzen 2002). However, these departures from
     standard A0 stars, as well as standard model atmospheres, are thought to only lead to minor
     modifications of its $uvby-$H$\beta$ indices (Paunzen et al. 2002). A more practical problem is
     that neither of the two CM66 transformation equations from H$'\beta$ to H$\beta$ will give a 
     fully satisfactory fit due to the fact that the index for Vega should be transformed 
     intermediately between the B star and the 
     A,\,F star sets. Several tests were performed which all pointed in the direction that 
     Vega should be transformed to the standard system via an equation somewhere in  
     between the B and the A,\,F transformations but with a heavier weight for the latter.
     By using all listed A0 stars in Table\,II of CM66, a special transformation between
     H$'\beta$ and H$\beta$ for A0 dwarf stars was established,
     as can be seen in Figure \ref{fig:betatransf}. 
     The vertical line in the plot represents the observed H$\beta$ value for Vega (Hauck \& 
     Mermilliod, 1998).   
     However, the observed H$'\beta$ value of Vega is not known.
     In order to determine the zero point c in the transformation from calculated 
     (H$''\beta$) to ``observed'' H$'\beta$ values, H$''\beta$\,=\,H$'\beta$\,+\,c, 
     we have therefore adopted an H$'\beta$ value for Vega derived from the A0 stars line
     in Figure \ref{fig:betatransf}, read off at the observed H$\beta$ for the star. 
     After correction of the model H$''\beta$ value to H$'\beta$ 
     we have then calculated the model H$\beta$ values by using the empirical 
     A,F-transformation relation as mentioned earlier.\\

      The transmission functions of the filters used by Olsen (1983, 1984)
      and Schuster \& Nissen (1988) depart from those on which the system 
      was originally based (Crawford 1966). Filter profiles representative for these
      newer studies are given by Helt et al. (1987) and Bessell (2005). In
      particular, the more recent
      $v$ band is narrower, and its effective wavelength shifted by
      about  25\,\AA\ towards the red. To test the effects of this, we have
      calculated colours with the Helt et al. (1987) profiles as an
      alternative to the Crawford set, and then applied the transformations as 
      given by Schuster \& Nissen (1988) back to the standard system to mimic 
      the procedure of the observers. From this, we find changes of
      the calculated values, amounting to typically 0.02 magnitudes or less
      in the $m_{1}$ and $c_{1}$ index values and only 0.002 in $(b-y)$. 
      In particular, the changes in the
      differential values measuring the sensitivity of $m_1$ to metallicity
      and of $c_1$ to gravity are small,
      in $\delta m_1/\delta$[Me/H] typically 0.003 and in $\delta c_1/\delta
      \log g$ typically 0.01. Such effects do not change our 
      conclusions in the present paper.

      \subsection{Model colours}
      The model colours, with zero-point added using Vega observations, are 
      supplemented the present paper electronically.

\section{Comparison star samples}
 \label{sec:stellsamp}
   In order to test the reliability of our calculated colours, we selected a sample
   of standard stars. 
   These were taken from various sources: 
   one subset with well determined spectroscopic parameters 
   was selected from The Bright Star Catalogue (Hoffleit \& Warren 1995), another from 
   the $uvby$ standard stars listed by Crawford \& Barnes (1970), a third from
   the list of metal-poor stars in Schuster \& Nissen (1988), a fourth among stars that have been 
   observed by the Hubble-STIS spectrograph and a fifth from the stars listed in the study of Pop 
   II stars by Jonsell et al. (2005). Altogether 388 stars were thus selected. The fundamental 
   parameters were taken from the sources listed, or from other sources given in the SIMBAD 
   catalogue and judged to have high quality. Complementary photometry 
   was also obtained from SIMBAD. Parameter determinations based on $uvby-$H$\beta$ 
   photometry were avoided as far as possible, since we aimed at testing calibrations based on 
   this photometric system relative to parameters 
   based on more fundamental methods. In practice, this usually means 
   effective temperatures based on the infrared-flux method and gravities and metallicities based
   on high-resolution spectroscopy. 
   The effective temperatures gathered from the Jonsell et al (2005) sample are calculated
   with the Alonso et al. (1996) calibration. These stars however, constitute less than 5\,\% of our
   total sample. In the case of multiple sources, i.e., stellar values listed in more than one 
   of our selected catalogues, a mean value was used. 
   The $uvby$ indices of the standard stars were dereddened by means of the algorithm and 
   computer code of Hakkila (1997). For most of the stars, in particularly the dwarfs,
   $E$(B-V) was smaller than $\sim$0.01.

\begin{figure}[hbtp]
      \resizebox{\hsize}{!}{\includegraphics[angle=90]{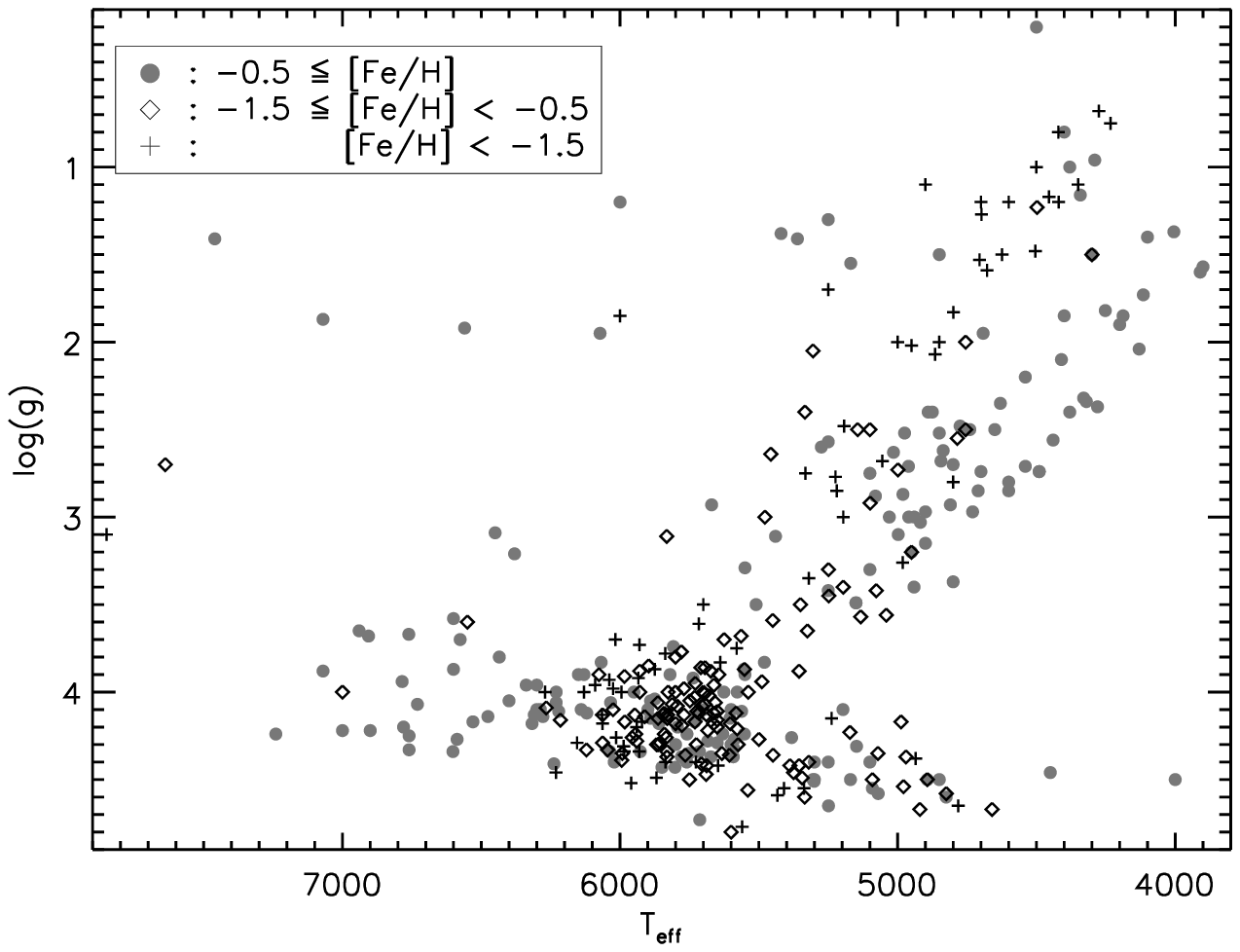}}
      \caption{Fundamental parameters, i.e. $T_{\rm eff}$, [Fe/H] and $\log g$,
               for the standard sample.}
      \label{fig:comp_stars}
     \end{figure}

   Altogether, the standard stars span a volume in the parameter space ranging from 3900\,K to 
   7850\,K in $T_{\rm eff}$, 0.20 to 4.80 in $\log g$ (cgs units) and --3.0 dex to 0.45 dex in 
   [Fe/H]. Data for the full standard sample is accessible 
   electronically from A\&A as on-line material supplementing the present paper.   
   As seen in Figure \ref{fig:comp_stars} our standard sample satisfactorily covers the fundamental
   parameter space, although one would need some more cool ($T_{\rm eff} < 5000$) and hot 
   ($T_{\rm eff} > 6300$) dwarf stars ($\log g > 4.0$). For this purpose two complementary 
   homogeneous samples were tested; the stellar sample published by Casagrande et al (2006, 
   hereafter C06) and Valenti \& Fischer (2005, hereafter VF05).
   $uvby$ colour indices for both samples were taken from Hauck \& Mermilliod (1998).
   These complimentary stellar samples cover the parameter spaces: $4700 \leq T_{\rm eff} \leq 6600$, 
   $3.1 \leq \log g \leq 5.1$ and $-1.9 \leq \rm [Fe/H] \leq 0.5$  for the 
   VF05 sample and
   $4406 \leq T_{\rm eff} \leq 6556 $, $\log g \sim4.5$ and $-1.87 \leq \rm [Fe/H] \leq 0.34$ for the
   C06 sample.
   Despite that the lowest $\log g$ values in the VF05 sample are characteristic of giants/sub-giants, 
   a majority (93\,\%) of the VF05 stars are dwarfs with $\log g \geq 4.0$. In the C06 
   sample all stars are 
   assumed to be dwarfs with $\log g \sim4.5$. A majority (85\,\%) of the stars in the VF05 sample 
   are also metal-rich, [Fe/H] $\geq -0.2$, whereas the C06 sample is evenly 
   distributed in metallicity. Likewise, the effective temperatures of the C06 sample 
   (determined by the authors's new calibration of the infrared-flux method) are evenly 
   distributed but the VF05 sample is biased towards higher temperatures, $T_{\rm eff} 
   \geq 5500$ (74\,\%). Among the stars in the standard sample, 18 and 72 are found in 
   the C06 and VF05 samples, respectively.
   For the stars in the VF05 sample no significant trends of discrepacies for the 
   values given, $T_{\rm eff}$, $\log g$ and [Fe/H], were found. In the C06 sample 
   we find an overall difference in given
   effective temperatures, growing for higher effective temperatures at 
   low metallicities ($T_{\rm eff\,standard} - T_{\rm eff\,Casagrande} \approx -175$\,K at 
   $T_{\rm eff} \sim6000$\,K and $-2.0 \leq $ [Fe/H] $\leq -1.0$).

\section{The effective-temperature calibration}
\label{sec:teffcalib}
   Precise determinations of effective temperatures of stars are critical for a
   variety of reasons, not only for direct applications such
   as comparisons of isochrone calculations to observed colour-magnitude diagrams,
   but also in more indirect applications e.g. determinations of chemical abundances. 
   Photometric indices play a critical role when determining the effective temperatures of 
   stars. For stars located on the main sequence, model atmosphere calibrations of these 
   indices may be particularly important since determinations of diameters are generally 
   few and poor.\\
   \indent
   The Str\"omgren $(b-y)$ index provides a sensitive 
   temperature measure for F, G and K stars. The H$\beta$ index 
   is another frequently used criterion for deriving temperatures. 
   We study both indices here.
   
\begin{figure*}[hbtp]
     \begin{center}
       \includegraphics[width=13cm,angle=90]{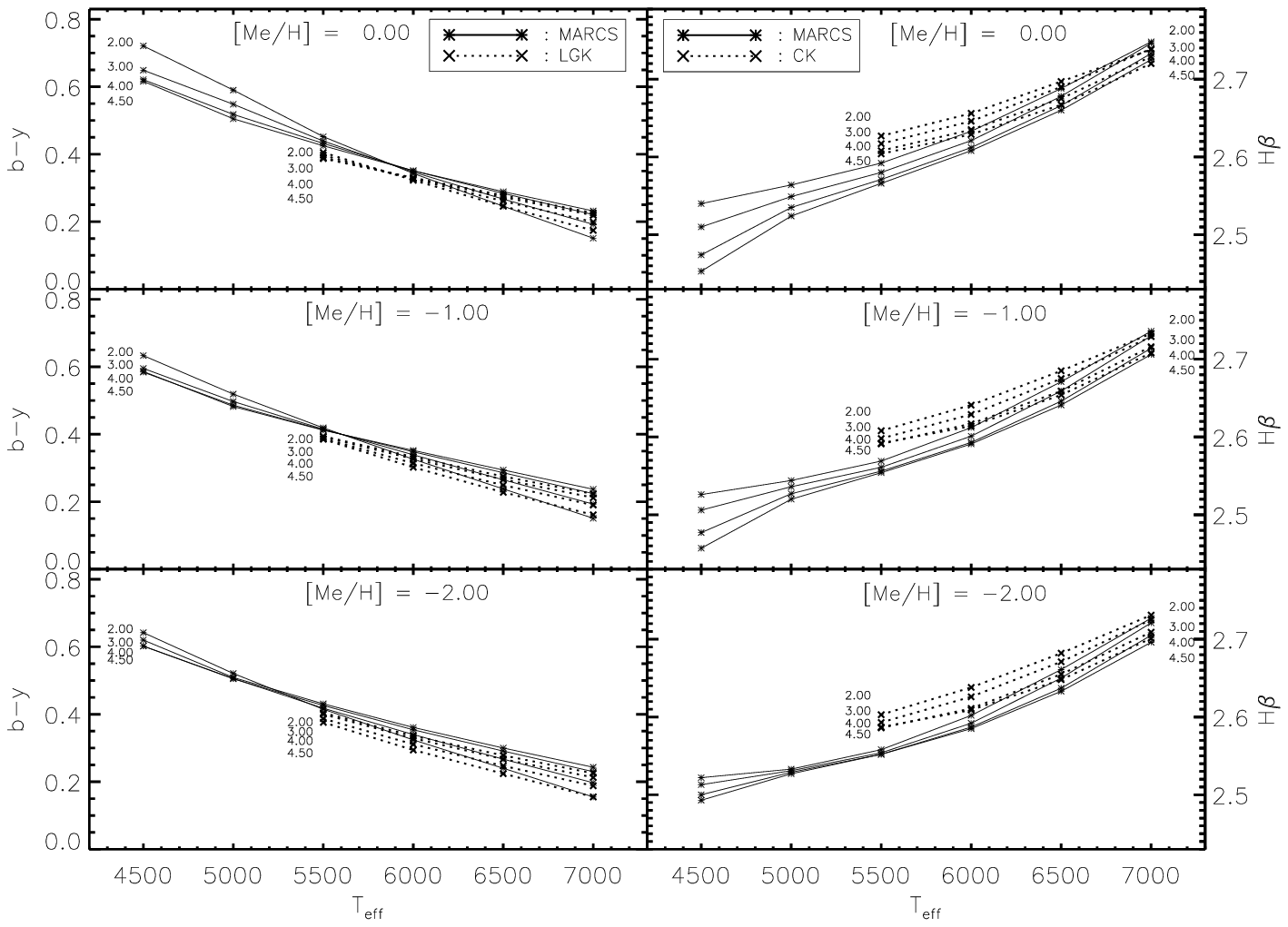}
       \caption{The $(b-y)$ colour index (left) and H$\beta$ index (right) versus 
                effective temperature for 
               MARCS and Kurucz (Lester, Gray \& Kurucz (1986), LGK, Castelli \& Kurucz, 2006, CK) 
               model atmospheres with different [Me/H] and $\log g$ 
               (indicated to the left and right of each curve). Corresponding
               theoretical MARCS $(b-y)$ and H$\beta$ curves for [Me/H] = 0.5, --0.5, and --3.0 
               can be seen in 
               Figure \ref{fig:byteffwithstars} and Figure \ref{fig:Hbetateffwithstars}, respectivley.}
      \label{fig:bybetateffteor}
      \end{center}
     \end{figure*}

\begin{figure*}[hbtp]
     \begin{center}
      \includegraphics[width=13cm,angle=90]{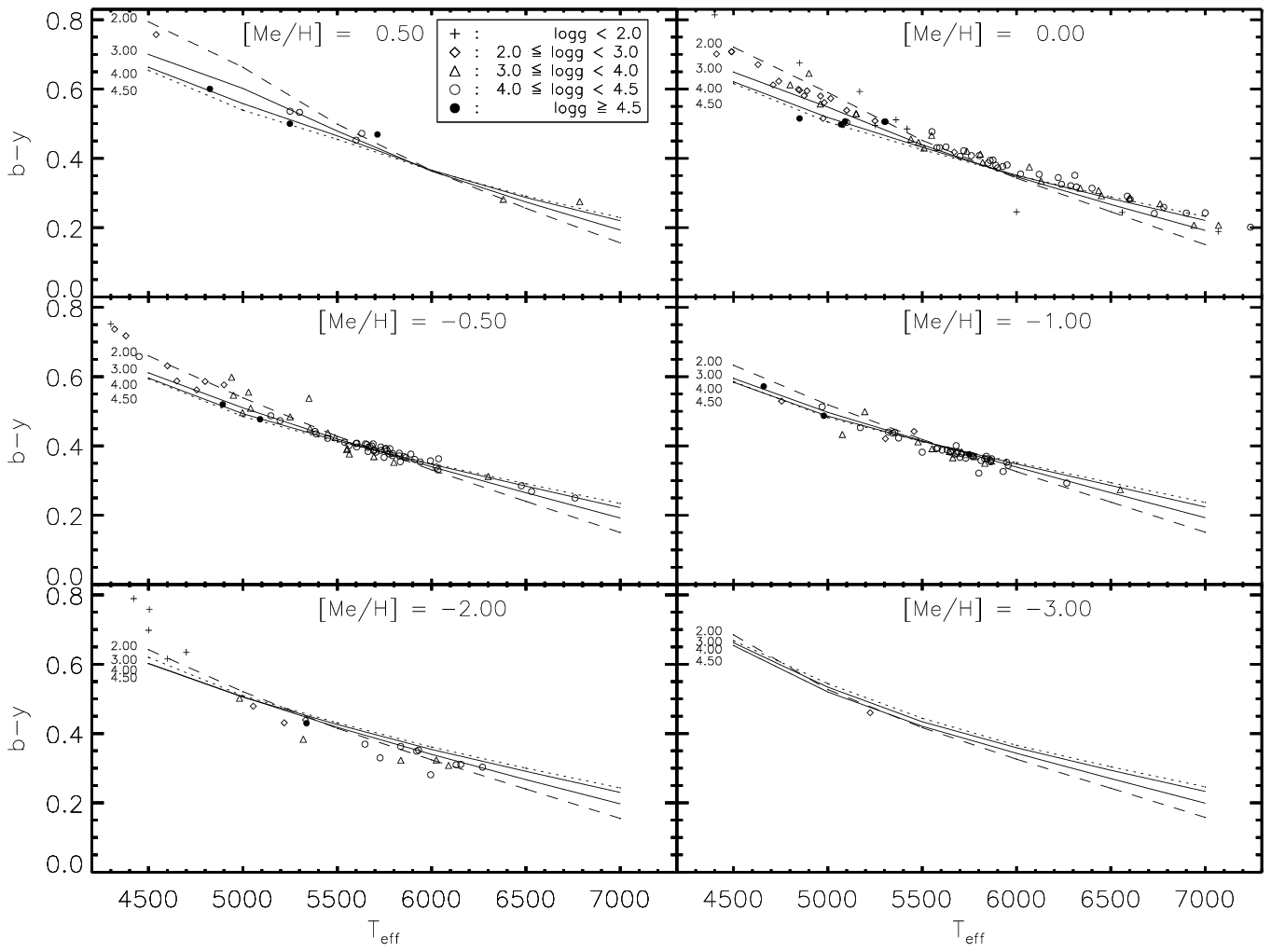}
      \caption{Theoretical $(b-y)$ values versus effective temperature for models with 
        different [Me/H] and $\log g$ plotted together with values for the standard stars
        (with [Fe/H] within $\pm$0.1\,dex of the given [Me/H]).
        To guide the eye the theoretical $\log g$ curves for 2.0 and 4.5 are plotted as 
        a dashed and a dotted line, respectively.}
      \label{fig:byteffwithstars}
     \end{center}
     \end{figure*}

\begin{figure*}[hbtp]
     \begin{center}
      \includegraphics[width=13cm,angle=90]{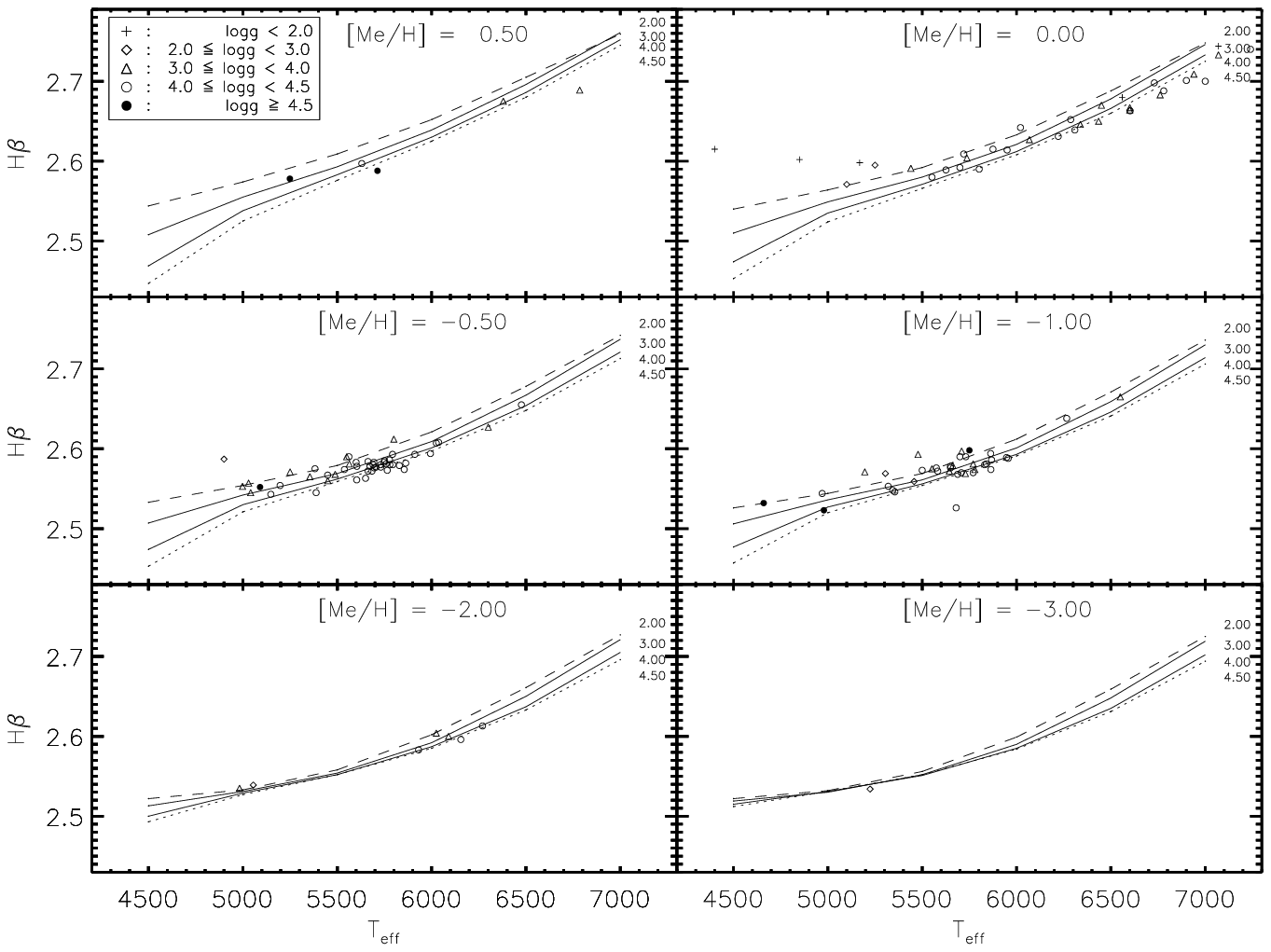}
      \caption{Theoretical H$\beta$ values versus effective temperature for models with 
               different [Me/H] and $\log g$ plotted together with values for the 
               standard stars (with [Fe/H] within $\pm$0.1\,dex of the given [Me/H]).
               To guide the eye the theoretical $\log g$ curves for 2.0 
               and 4.5 are plotted as a dashed and a dotted line, respectively.}
      \label{fig:Hbetateffwithstars}
      \end{center}
     \end{figure*}

   In Figure \ref{fig:bybetateffteor} (left) we explore the temperature sensitivity of the 
   theoretical $(b-y)$ index for different metallicities.
   As a comparison we also plot $(b-y)$ for a set of Kurucz models (Lester, Gray \& 
   Kurucz, 1986). We note that the indices, calculated with the different model sets,
   more or less show the same behaviour although the models of Lester et al. are
   systematically bluer at given $T_{\rm eff}$.
   In comparison with observed values for stars our synthetic indices show a satisfactory 
   temperature sensitivity over the full temperature range (4500--7000\,K), see 
   Figure \ref{fig:byteffwithstars}.\\
   \indent
   In Figure \ref{fig:bybetateffteor} (right) the H$\beta$ index is plotted versus
   effective temperature, together with the theoretical data of Castelli \& Kurucz (2006).
   For the warmer part of the temperature range we find no differences in sensitivity,
   whereas for the cooler temperatures (5500\,K) we note 
   that MARCS indices have a steeper gradient, thus lower H$\beta$ values than those of 
   Castelli \& Kurucz. Our steeper gradient tends to agree with observed stellar colours, Figure 
   \ref{fig:Hbetateffwithstars}, which is probably due to improvements of the broadening
   theory of H$\beta$.

 \subsection{$T_{\rm eff}(b-y)$}

  \label{sec:teffby}
    Several attempts (Alonso et al. 1996, Nordstr\"om et al. 2004, Ram\'{i}rez \& 
    Mel\'{e}ndez 2005, Holmberg et al. 2007) 
    have been made to derive empirical calibrations for $uvby-$H$\beta$ colours,
    where the fundamental parameters are expressed as explicit formulae in the photometric 
    indices. We will here derive such calibration equations by using theoretical colours.
    We make use of the formal expressions of Alonso et al. 
    (1996, 1999, hereafter A96 resp. A99) based on $(b-y)$ 
    (their eq. 9 and 14 \& 15, respectively) but derive new coefficients via an iterative
    least squares method and by using the model colours.\\
    \indent
    The result for dwarfs ($\log g \geq 4.0$) are given numerically
    in Appendix \ref{sec:app_teffByBetaCalib} and can be seen in 
    Figure \ref{fig:bydwarf} where we also merge our standard sample with the C06 sample
    and plot temperatures for the individual stars derived from our theoretical calibration 
    equation.  
    The standard deviations for our calculated temperatures with respect 
    to literature values can be found in Table \ref{tab:standDev}
    A linear regression for the calculated temperatures of the two merged samples 
    is also shown in Figure \ref{fig:bydwarf} and yields a dispersion 
    $\sigma$ = 117\,K. We note that our theoretical calibration gives higher
    temperatures, as compared with literature values, by $\sim$100\,K for the cooler stars 
    ($T_{\rm eff}$$\,<\,$5000\,K) and tends to lead to lower temperatures ($\sim$100--150\,K) 
    than the values found in the literature for hotter stars ($T_{\rm eff}$$\,>\,$6500\,K).
    Our calibration suggests lower temperatures than the 
    values listed in VF05 by 100-200\,K for stars within the range of 5200-6300\,K.
    Below that our theoretical calibration gives systematically $\sim$100\,K higher temperatures
    for the VF05 stars, in accordance with the tendencies found also for the other comparison
    samples. The standard deviation of the calculated temperatures for the
    VF05 sample can be seen in Table \ref{tab:standDev}.
    
    It is also interesting to compare our results to previous empirical calibrations. 
    The standard deviations of the calculated temperatures using four different 
    empirical calibrations applied on our three comparison samples, are
    listed in Table \ref{tab:standDev}. In order to illustrate the empirical trends the resultant 
    linear regressions for the merged standard and C06 samples are shown in Figure \ref{fig:bydwarf}
    (note the shifted scale by 1000\,K).
    \begin{table}
     \caption{Standard deviations for $T_{\rm eff}$ and [Fe/H] derived
              using models as compared with literature values of the different
              comparison samples.}
     \begin{tabular}{c|ccc||cc}
     \hline \hline
                   &   & \textbf{$T_{\rm eff}$} &  & \textbf{[Fe/H]} &  \\
                   & Dwarfs  &          & Giants  & G-stars & F-stars  \\
     \textbf{Theoretical} & $(b-y)$ & H$\beta$ & $(b-y)$ &         &   \\  
     \hline	                                                       
         Standard  & 133     &  151     & 144     &  0.263  & 0.254    \\
         C06       & 101     &  ---     & ---     &  0.321  & 0.190    \\
         VF05      & 185     &  ---     & ---     &  0.392  & 0.399    \\
         All       & 172     &  ---     & ---     &  0.368  & 0.355    \\
      \hline \hline
      \textbf{A06,99\,+\,S\&N89} & A06  &   &  A99 &  S\&N89   &          \\
     \hline	                                      
         Standard  & 110     & 156      & 116     &  0.183  & 0.182    \\
         C06       & 118     & ---      & ---     &  0.158  & 0.128    \\
         VF05      & 135     & ---      & ---     &  0.128  & 0.146    \\
         All       & 130     & ---      & ---     &  0.141  & 0.157    \\
      \hline \hline                            
      \textbf{HNA07} &       &          &         &         &          \\
     \hline	       
         Standard  & 101     & ---      & ---     &  0.208  & 0.164    \\
         C06       & 92      & ---      & ---     &  0.159  & 0.132    \\
         VF05      & 101     & ---      & ---     &  0.123  & 0.130    \\
         All       & 100     & ---      & ---     &  0.144  & 0.140    \\
     \hline \hline                               
     \textbf{RM05} &         &          &         &         &          \\
     \hline	                
         Standard  & 155     & ---      & 105     &  0.268  &  0.176   \\
         C06       & 110     & ---      & ---     &  0.176  &  0.139   \\
         VF05      & 128     & ---      & ---     &  0.116  &  0.081   \\
         All       & 134     & ---      & ---     &  0.161  &  0.122   \\
      \hline
     \end{tabular}
     \label{tab:standDev}
    \end{table}

    The calibration results for giants ($1.5 \le \log g \le 3.5$) is presented in Appendix 
    \ref{sec:app_teffByCalibGiant} and shown in Figure \ref{fig:bygiant}. The standard deviations 
    with respect to effective temperatures of the standard stars are shown in Table 
    \ref{tab:standDev} together with the standard deviations for the same sample when
    using the A99 empirical calibration.

 \subsection{$T_{\rm eff}$(H$\beta$)} 
\begin{figure}[hbtp]
      \resizebox{\hsize}{!}{\includegraphics[angle=90]{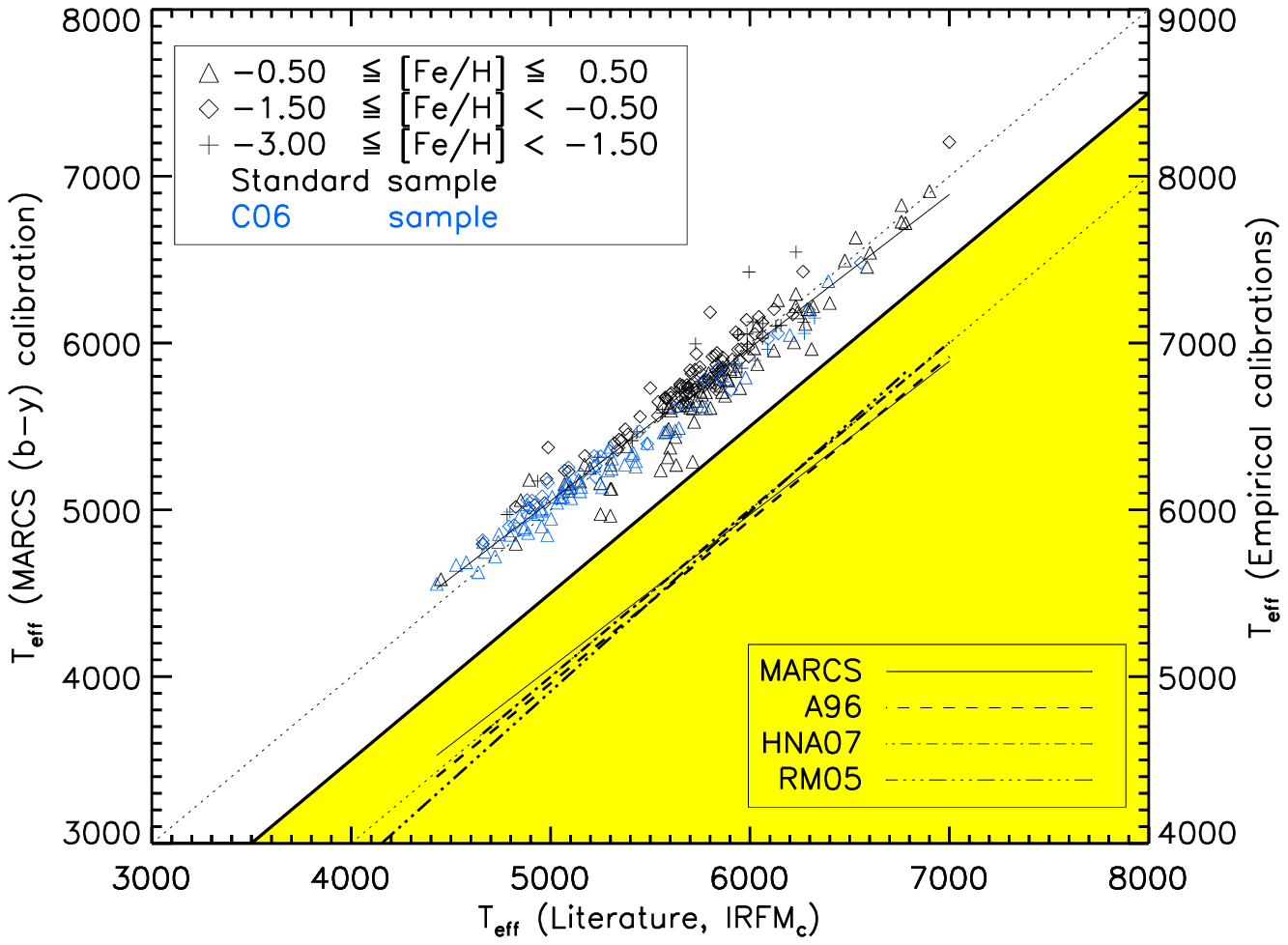}}
      \caption{Comparison of calibrations of $(b-y)$ for dwarf stars with individual values 
      for the standard stars and C06 plotted. The solid line represents a fit of the effective 
      temperatures adopted for the stars relative to the corresponding values 
      obtained from the theoretical calibration. Below that (shaded area and right y-axis) 
      corresponding linear regressions of the adopted effective temperatures relative to 
      empirical calibrations (Alonso et al. 1996, Holmberg et al. 2007 and Ram\'{i}rez \& 
      Mel\'{e}ndez 2005) are shown. The dotted line is a one-to-one line.}
      \label{fig:bydwarf}
     \end{figure}

\begin{figure}[hbtp]
      \resizebox{\hsize}{!}{\includegraphics[angle=90]{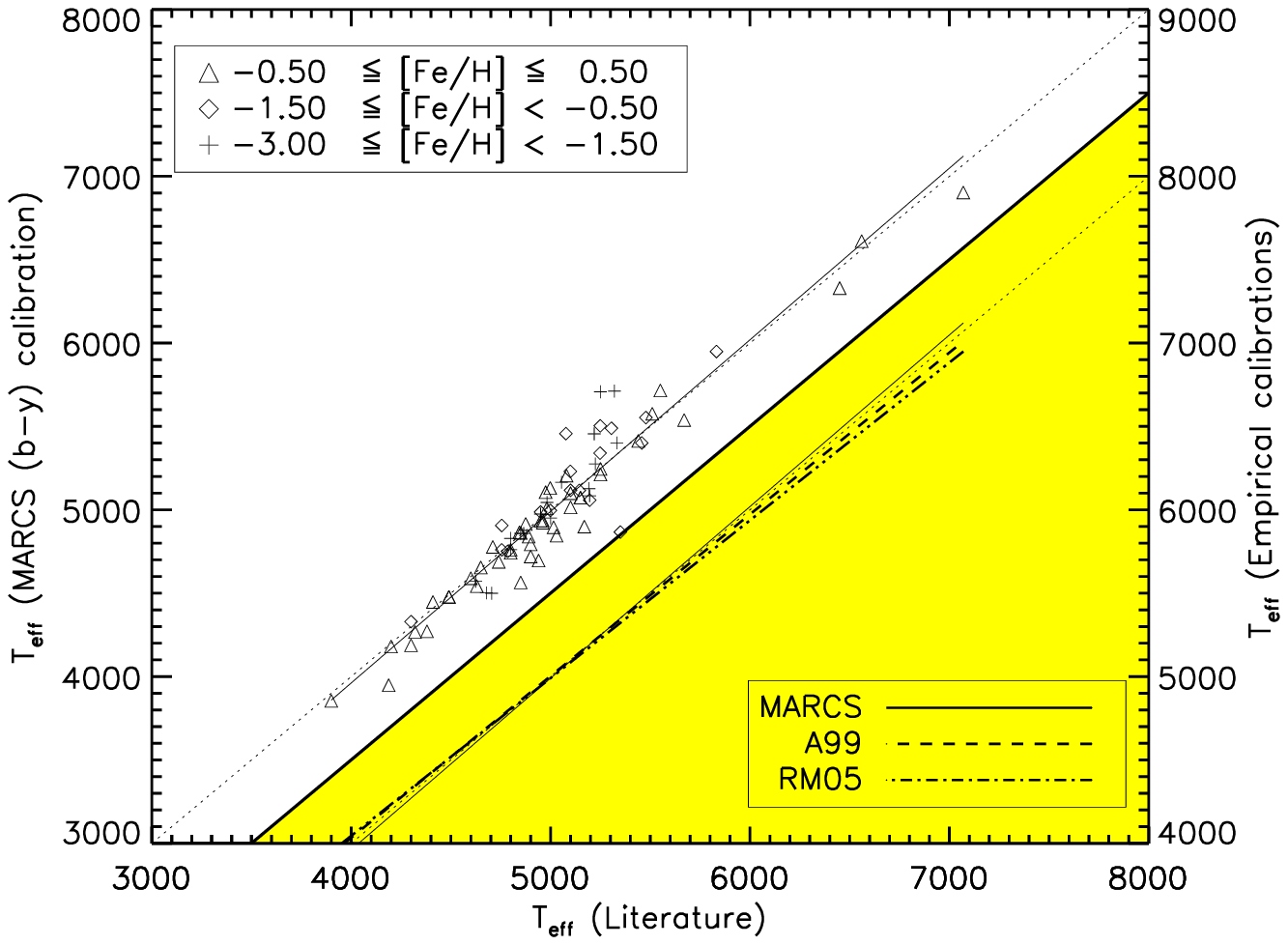}}
      \caption{Comparison of calibrations of $(b-y)$ for giant stars with individual values for 
      the standard stars plotted. The solid line represents a fit of the effective temperatures
      adopted for the standard stars relative to the corresponding values obtained from the 
      theoretical calibration. Below that (shaded area and right y-axis) 
      corresponding linear regressions of the adopted effective temperatures relative to 
      empirical calibrations (Alonso et al. 1999 and Ram\'{i}rez \& Mel\'{e}ndez 2005) are shown.  
      The dotted line is a one-to-one line.}
      \label{fig:bygiant}
     \end{figure}

\begin{figure}[hbtp]
      \resizebox{\hsize}{!}{\includegraphics[angle=90]{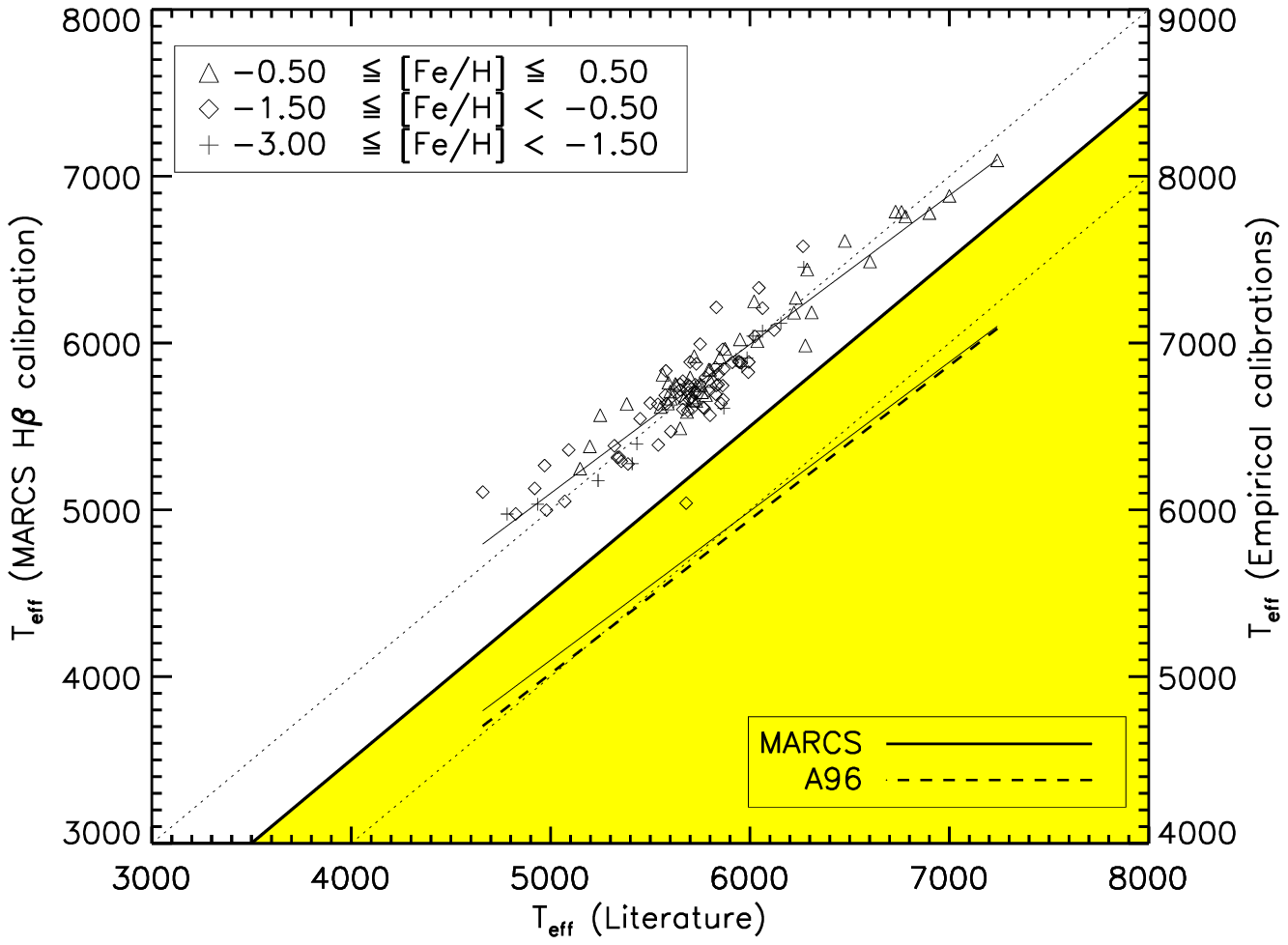}}
      \caption{Comparison of calibrations of $T_{\rm eff}({\rm H}\beta)$ for 
      dwarf stars with individual values for the standard stars
      plotted. The solid line represents a fit of the effective temperatures
      adopted for the standard stars relative to the corresponding values obtained from the 
      theoretical calibration. Below that (shaded area and right y-axis) a
      corresponding linear regression of the adopted effective temperautres relative to an
      empirical calibration (Alonso et al. 1996) is shown. Note that the star located in the
      shaded area belongs to the theoretical calibration and should therefore be read off at the
      left y-axis. The dotted line is a one-to-one line.}
      \label{fig:betadwarf}
     \end{figure}

   A theoretical temperature calibration for dwarfs
    was also derived for the H$\beta$ index, with a formal expression adopted 
    from A96, Eq. 10, see Appendix \ref{sec:app_teffByBetaCalib}. The result can be 
    seen in Figure \ref{fig:betadwarf}.
    Standard deviations with respect to the literature values of the standard sample
    are listed in Table \ref{tab:standDev}, values somewhat higher than
    the calibration based on $(b-y)$.  
    Here we are, however, using different stars in the 
    $T_{\rm eff}$$(b-y)$ and $T_{\rm eff}$(H$\beta$) equations (fit limits defined in A96). 
    To make a fair comparison between the 
    two calibrations we have restricted our sets of stars to be identical for both calibrations. 
    Such a comparison is made in Sec. \ref{sec:hbvsby}. 
    A linear regression for the calculated effective temperatures is also shown in Figure 
    \ref{fig:betadwarf}. The dispersion around this fit is $\sigma$ = 143\,K.
    We see that our theoretical calibration suggests somewhat higher effective temperatures, 
    of roughly 100\,K, in the cooler part of the temperature range (T$_{\rm eff}$\,$\sim$\,5000\,K),
    and that it possibly implies lower effective temperatures of the order of $\sim$\,100\,K
    for the warmest stars ($T_{\rm eff}$\,$>$\,6500\,K). Some of these tendencies may possibly
    be traced also in the $(b-y)-T_{\rm eff}$ calibration (c.f. Fig. \ref{fig:bydwarf}).
    As an alternative, the effective temperatures for the standard sample are calculated
    with the empirical calibration equation by A06. The standard deviations with respect
    to literature values are presented in Table \ref{tab:standDev}, and a linear regression
    to the calculated effective temperatures is also plotted in Figure \ref{fig:betadwarf}.

 \subsection{$T_{\rm eff}$($b-y$) vs $T_{\rm eff}$(H$\beta$)}
  \label{sec:hbvsby}
\begin{figure}[hbtp]
     \resizebox{\hsize}{!}{\includegraphics[angle=90]{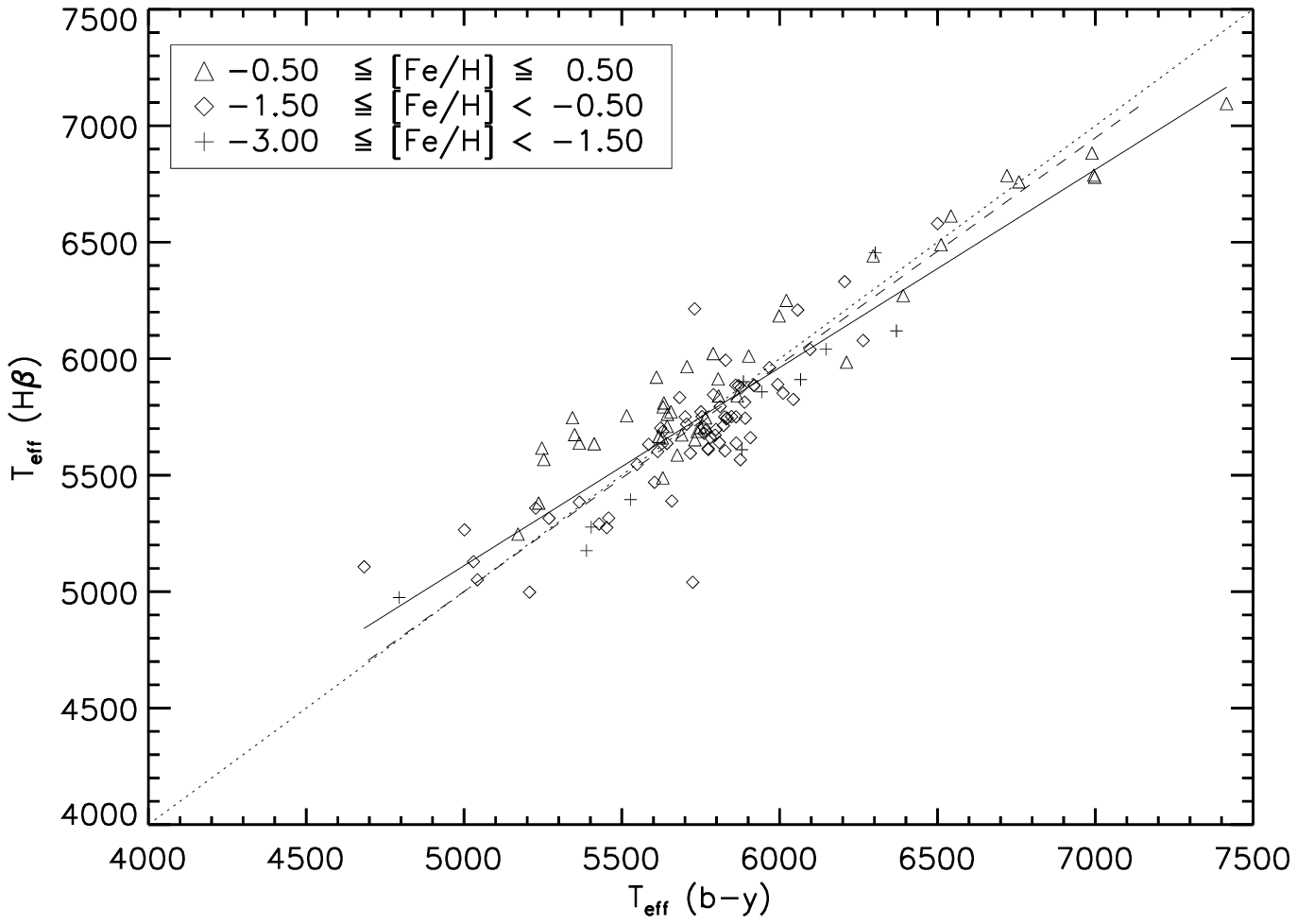}}
     \caption{The effective temperatures calculated by the equations based on H$\beta$ 
              and $(b-y)$. Solid line theoretical 
              fit, dashed line empirical fit (Alonso et al. 1996). 
              The dotted line is the one-to-one relation.}
     \label{fig:byvsbeta}
    \end{figure}

    Now we compare the derived temperatures based on $(b-y)$ and H$\beta$,
    respectively. 119 stars matching the restrictions in the parameter space for 
    both calibrations were
    selected out of our standard sample. The effective temperatures
    based on $(b-y)$ and H$\beta$ and the result are displayed in Figure \ref{fig:byvsbeta}. 
    The test reveals that there is some disagreement between the two theoretical 
    calibrations. 
    The H$\beta$ based equation suggests higher temperatures for
    the cooler stars ($T_{\rm eff}$\,$\lesssim$\,5000\,K) and lower temperatures for
    the warmer stars ($T_{\rm eff}$\,$<$\,6900\,K) of some 200\,K.
    The empirical equation of Alonso et al. (1996) shows more or less the same spread 
    for the standard sample but no 
    significant departures from the 1-1 line, which is also to be expected. The failure for the 
    theoretical calibration
    for the coolest stars is not very remarkable, since the metal lines for those 
    stars are strong and dominate the H$\beta$ line, and other parameters affect the
    index such as gravity. The departure in the hotter end is
    only dependent on a few standard stars.\\ 
    \indent
    It is worth noting that the difference between the $(b-y)$ and H$\beta$ calibrations
    responds in opposite directions to different gravities. 
    I.e., the temperature sensitivity 
    $\delta(index)/\delta(T_{\rm eff})$ decreases with increasing gravity for $(b-y)$ 
    while it increases for $T_{\rm eff} \le 5000\,K$ for H$\beta$ 
    (see Fig. \ref{fig:bybetateffteor}). 
    As is seen, these differences are also metallicity dependent, and tend to 
    vanish for small metallicities. However, for stars with known reddening and assuming 
    e.g. the metallicity to be known and relatively large, this may make it possible to 
    obtain a temperature and rough gravity classification from $(b-y)$ and H$\beta$, only.
   
\section{The metallicity calibration}
 \label{sec:metcalib}
   The basic aim of $m_1 \equiv (v-b)-(b-y)$ was to measure the total intensity of the 
   metal lines in the $v$ band. As was early appreciated by Str\"omgren, for late F and G 
   stars of Pop I these lines are, however, to a large extent located on the flat part of the 
   Curve of Growth and are thus not very sensitive to metallicity, but rather to microturbulence. 
   Moreover, for the hotter stars, the H$\delta$ line is strongly affecting the band, and for 
   stars later than G5, CN lines of the (0,1) band in the Violet System are also significant. 
   Thus, the effects of the value of the mircroturbulence parameter, as well as of the 
   individual CNO abundances, e.g. due to dredge-up of CNO processed material 
   from the interior, must be taken into consideration. Certainly, $m_{1}$ also varies with 
   effective temperature and, to a less degree, with surface gravity. \\
  
\begin{figure*}[hbtp]
     \begin{center}
     \includegraphics[width=13cm,angle=90]{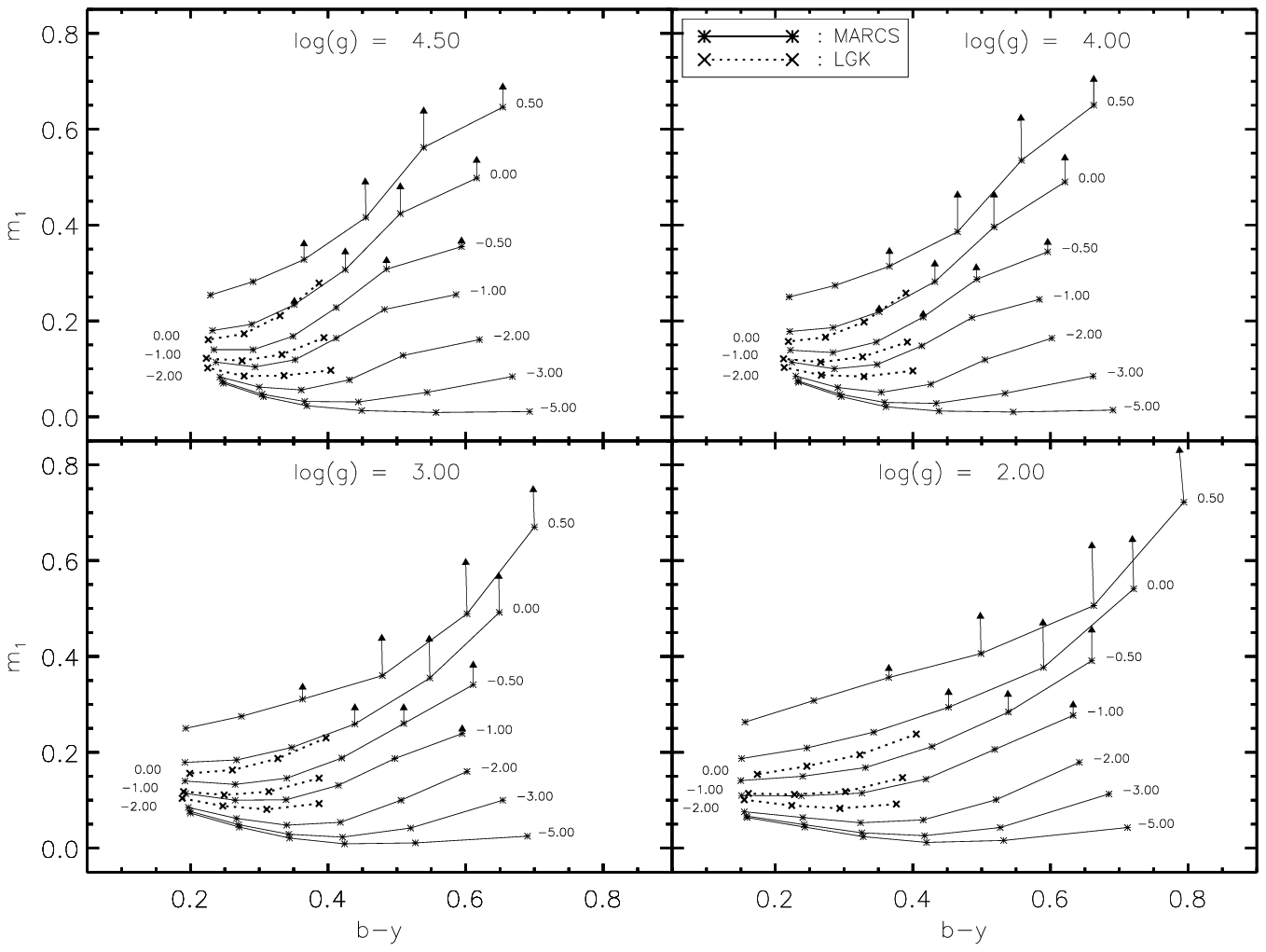}
     \caption{The $m_{1}$ index versus $(b-y)$ for MARCS and Lester, Gray \& Kurucz (1986, LGK) 
              model atmospheres. Arrows show the effect of increasing the N 
              abundances by a factor of 3.}
     \label{fig:m1vsby}
    \end{center}
    \end{figure*}

    The variation of the calculated $m_1$ with $(b-y)$ and metallicity 
    for the model atmospheres is shown in Figure \ref{fig:m1vsby} and
    compared with calculated indices by Lester et al. (1986). 
    As is seen, the index offers a good discrimination in metallicity except for 
    stars of Extreme Population II 
    for which it only works for the cooler end of the temperature interval. A 
    characteristic measure of the sensitivity of the index to overall metallicity is 
    $(\delta(m_1)/\delta$[Me/H]$)_{(b-y)}$, where the subscript denotes that the sensitivity
    is measured at a constant $(b-y)$. This quantity, as measured for models with --3.0 
    $\leq$ [Me/H] $\leq 0.5$ and $\log g$ $\approx$\,4.5 \& 4.0, is given in Table\, 
    \ref{table:m1sensteff}.   

    We may compare the sensitivity of the $m_{1}$ index with the empirical
    results. However, recent calibrations of the Str\"omgen photometry, e.g.
    by Holmberg et al. (2007), contain complex non-linear expressions in which all
    the indices are involved -- in principle a reasonable approach since e.g. also
    the $c_1$ index carries information on the metallicity. Since this
    metallicity dependence is, however, far from independent of that
    carried by the $m_1$ index the terms in the calibration expressions
    involving $c_{1}$ may well mask some of the dependence of $m_{1}$ on
    metallicity. So, as we wish to understand the way $m_{1}$ changes with
    metallicity, we have instead turned back to the earlier empirical calibrations
    like that of Nissen (1988), where the metallicity dependence of
    $m_{1}$ was still treated separately.
    
    Thus, Nissen (1988) finds
    \begin{eqnarray}
    \label{eqcalib1}
    \rm [Fe/H]=-(10.5+50(H\beta-2.626))\cdot \delta m_0 + 0.12
    \end{eqnarray}
    where $\delta m_0$ is calculated as $m_{stand}-m_1$ relative to the Hyades
    standard sequence at constant H$\beta$ and corrected for interstellar extinction.
    From Figure \ref{fig:m1vsby} or Table \ref{table:m1sensteff} we may estimate 
    $\delta m_1/\delta $[Me/H] at constant
    $(b-y)$. We can easily calculate the corresponding empirical
    quantity from the coefficient in Eq. (\ref{eqcalib1}) (correcting for the 
    derivative at constant H$\beta$ to constant $(b-y)$ by adding 0.03 which is 
    easily found with sufficient accuracy from the model colours) and then find 
    empirical sensitivities $\delta m_1/\delta $[Fe/H] of 0.093 (7000\,K) and 0.15 
    (5750\,K), with relevant effective temperatures within parentheses. These values 
    for the sensitivities are valid for [Fe/H]$\approx0.0$. The corresponding theoretical 
    sensitivities $\delta m_1/\delta $[Me/H] for dwarf stars are typically 0.10 and 0.14, 
    respectively. The agreement between the empirical
    and theoretical results is quite satisfactory. It seems possible that a
    basic reason why the theoretical calibration, in spite of this agreement, does
    not succeed very well for the more metal-poor stars (see below) may rather be due to
    the modelling of the fluxes of those stars than due to failures in
    calculating the change of flux with metallicity for solar-type stars. It may also be
    associated with the measured $v$ band being different to that computed, as a single
    transformation equation is unlikely to correct solar metallicity stars and metal-poor
    stars by the same amount.

\begin{figure*}[hbtp]
     \begin{center}
     \includegraphics[width=13cm,angle=90]{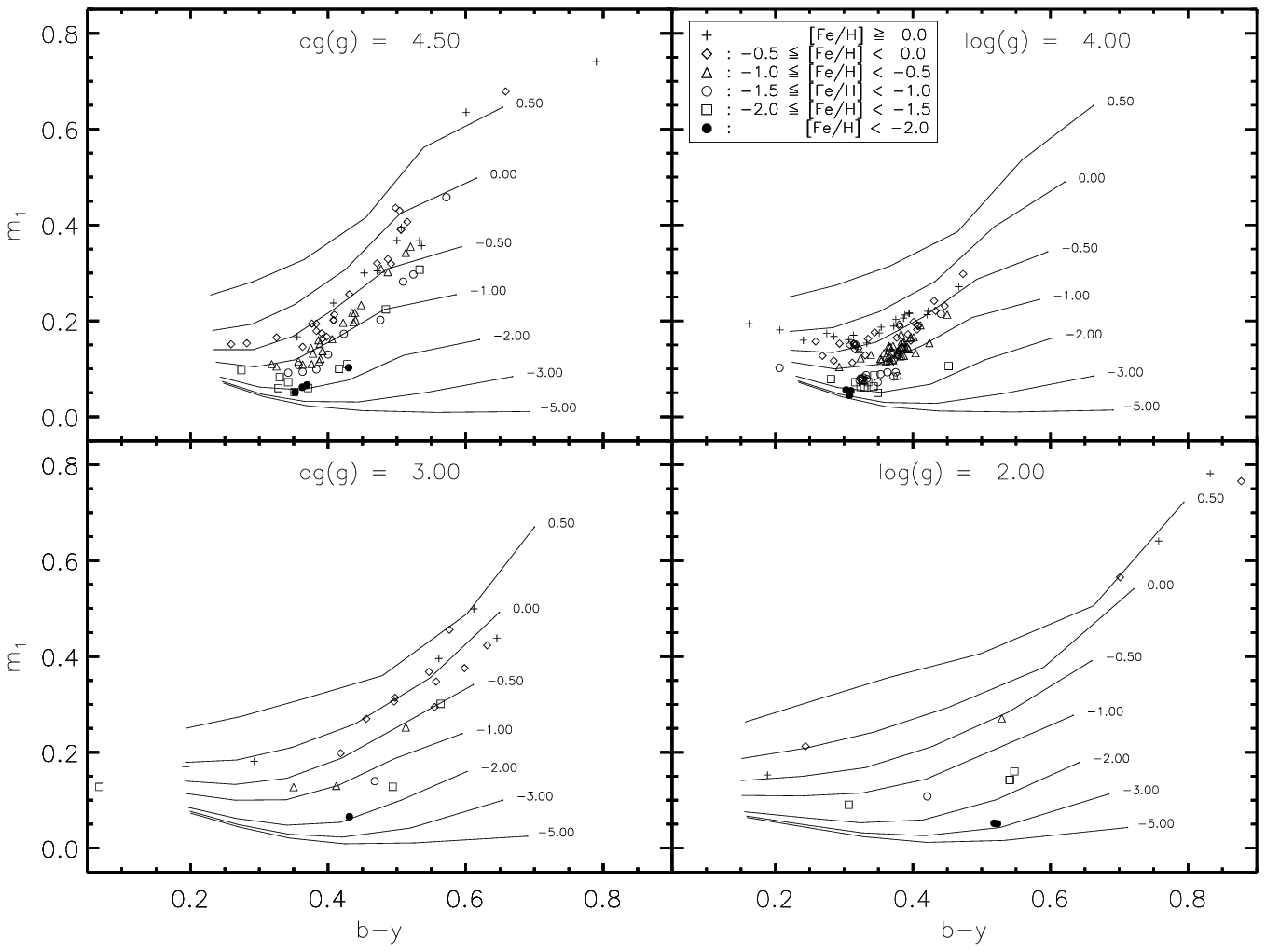}
     \caption{The m$_{1}$ index versus $(b-y)$ for MARCS models plotted together with
              standard stars of different gravities in different panels 
              ($\log g$ for the stars within $\pm$0.1\,dex of the given $\log g$). The 
              metallicities of the stars are indicated by different symbols.}
     \label{fig:m1vsbywithstars}
     \end{center}
    \end{figure*}

    In order to further explore the properties of the calculated $m_1$ indices 
    we have plotted individual stars with fairly well-determined fundamental parameters,
    chosen from our standard sample in the $m_1-(b-y)$ diagram. 
    As is seen in Figure \ref{fig:m1vsbywithstars}, these stars match the calculated indices 
    relatively well, although there seems to be a tendency of the sensitivity of the $m_1$ 
    index to metallicity to be exaggerated by the model fluxes for the hotter stars. Also, 
    it is clear from Figure \ref{fig:m1vsbywithstars} that the metal-rich stars lie somewhat 
    low, possibly suggesting that the zero-point of the $m_1$ index as determined from Vega 
    may be somewhat in error. 
    
    \begin{table}[h!]
    \caption{The metallicity sensitivity, $\delta(m_1)/\delta$[Me/H]$)_{(b-y)}$, for 
              models with $\log g$\,=\,4.5 \& 4.0.} 
     \label{table:m1sensteff}
     \begin{tabular}{cccc}
      \hline \hline
      $(b-y)$/[Me/H] & 0.5$-$($-$0.5) & $-$1$-$($-$2) & $-$2$-$($-$3)\\
      \hline 
      $\log g$\,=\,4.5 & & &        \\
      0.25 & 0.124 & 0.031 &  0.008 \\
      0.30 & 0.145 & 0.042 &  0.013 \\
      0.35 & 0.149 & 0.056 &  0.028 \\
      0.40 & 0.139 & 0.098 &  0.026 \\
      0.45 & 0.136 & 0.135 &  0.032 \\
      0.50 & 0.176 & 0.146 &  0.048 \\
      0.55 & 0.233 & 0.127 &  0.070 \\
      & & &                         \\
      $\log g$\,=\,4.0 & & &            \\    
      0.25 &  0.125  & 0.030  &  0.010 \\   
      0.30 &  0.143  & 0.041  &  0.014 \\  
      0.35 &  0.148  & 0.059  &  0.020 \\  
      0.40 &  0.137  & 0.081  &  0.031 \\  
      0.45 &  0.125  & 0.097  &  0.052 \\  
      0.50 &  0.145  & 0.100  &  0.076 \\  
      0.55 &  0.198  & 0.093  &  0.089 \\
      \hline
     \end{tabular}
    \end{table} 


    When studying the calculated indices of Lester, Gray \& Kurucz (1986) in Figure \ref{fig:m1vsby} 
    we find that they reproduce the observed sensitivity of the $m_1$ index more successfully 
    than ours. However, these
    authors have transformed their calculated colours to match a set of standard stars 
    and have thus scaled the amplitude of the $m_{1}$ index by a correction factor to 
    fit the observations. This is the probable reason for the closer agreement of those 
    calculations with observations. We note that our line list is more complete, and that 
    our treatment of the hydrogen line broadening (affecting H$\delta$ and thus the 
    $v$ band) is more accurate. 
    
    What is then the reason for our discrepancy for early F stars? We have compared the MARCS 
    model fluxes with observed solar and stellar fluxes from ground-based 
    and space observations (Edvardsson 2008, Edvardsson et al. 2008) and traced probably significant 
    discrepancies in the region 4000\,Å -- 5000\,Å, with empirical fluxes of the Sun and 
    solar-type stars being somewhat smaller than model fluxes in the $b$ band, while 
    the blue-violet fluxes from HST/STIS of the more metal-poor stars are clearly in excess 
    of the model fluxes in the violet-blue spectral region. These departures in both the $b$ 
    and $v$ band may conspire to cause the discrepancy in calculated metallicity sensitivity. 
    As discussed by Edvardsson et al. (2008), 3D model simulations suggest that these effects 
    may be due to thermal inhomogeneities in the stellar atmospheres. Other systematic 
    errors in the models, e.g. due to errors in opacities, line data and effects of 
    departures from LTE are probably less significant.
         
\begin{figure}[hbtp]
     \resizebox{\hsize}{!}{\includegraphics[angle=90]{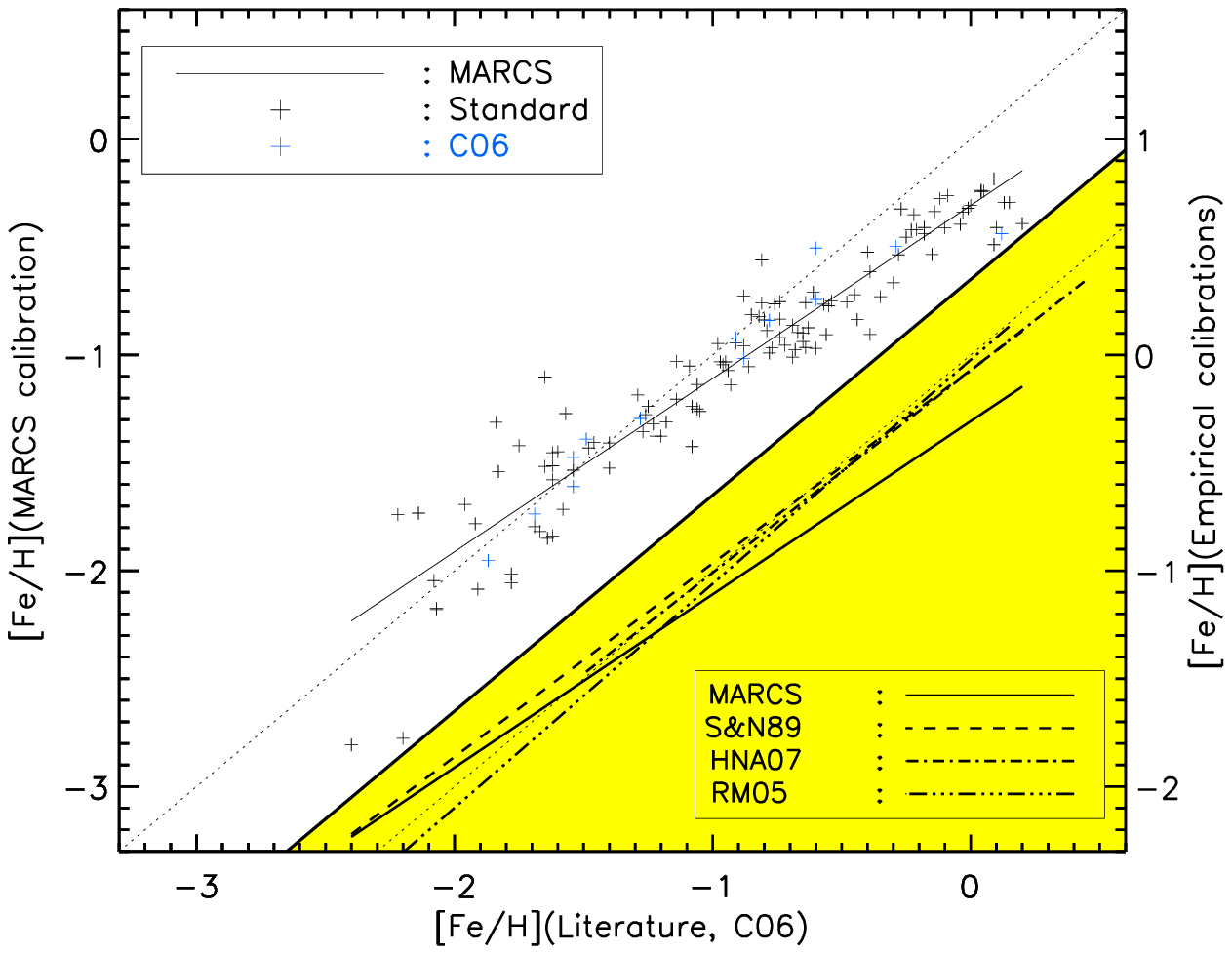}}
     \caption{The [Me/H] calibration for F stars from the standard sample and Casagrande
              et al. (2006, C06). The solid line represents a linear regression of the effective 
      temperatures adopted for the stars relative to the corresponding values 
      obtained from the theoretical calibration. Below that (shaded area and right y-axis) 
      corresponding linear regressions of the adopted effective temperautres relative to 
      empirical calibrations (Schuster \& Nissen 1989, Holmberg et al. 2007 and Ram\'{i}rez \& 
      Mel\'{e}ndez 2005) are shown. The dotted line is a one-to-one line.}
     \label{fig:fitF}
    \end{figure}

\begin{figure}[hbtp]
     \resizebox{\hsize}{!}{\includegraphics[angle=90]{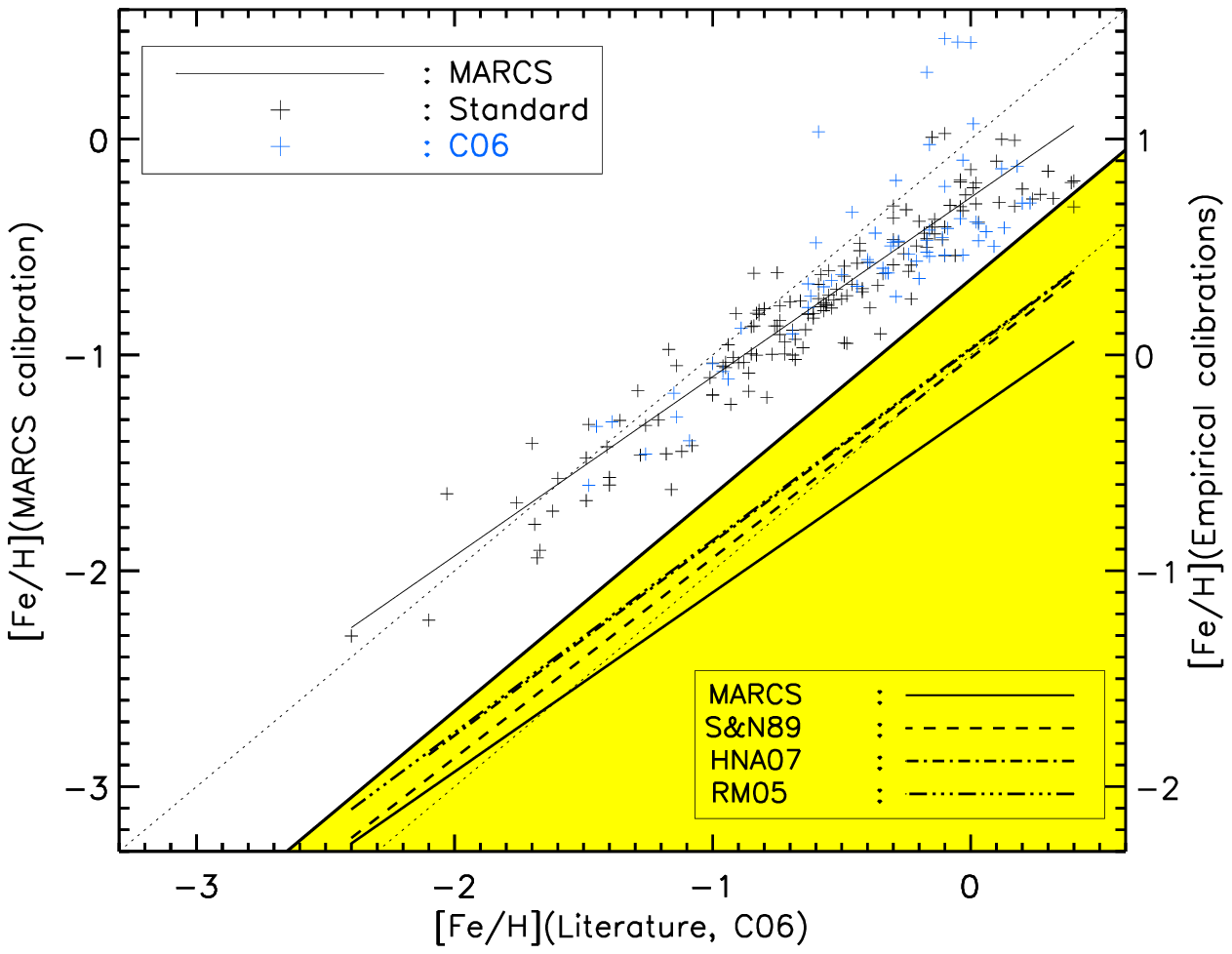}}
     \caption{The [Me/H] calibration for G stars. For plot description see Figure \ref{fig:fitF}.
              Note that empirical calibrations of HNA07 and RM05 overlap and could therefore
              be difficult to distinguish.}
     \label{fig:fitG}
    \end{figure}

    As a further test of our $m_{1}$ indices we will now be guided by the
    separate metallicity calibrations for F and G stars, respectively, by Schuster \& Nissen 
    (1989, eq. 2 \& 3, hereafter S\&N89). 
    The derived F-star equation is based on the $m_{1}$ and $(b-y)$ index and for the 
    G-star equation the $c_{1}$ index is also included. We derive a calibration expression 
    based upon {\it the form} of these equations and using our theoretical colour grid. 
    The results, when applying this to the standard stars and the C06 sample, can be seen in
    Figures \ref{fig:fitF} and \ref{fig:fitG}. 
    For F-stars we find standard deviations of derived [Me/H] values compared with literature values for the 
    standard-star sample as given in Table \ref{tab:standDev}.
    We note, that for the standard sample, our theoretical calibration tends to suggest lower metallicities than
    the adopted values for a majority of the stars in the sample and increasing differences with 
    increasing metallicity so that the differences amount to typically 0.3\,dex at
    [Me/H]\,=\,0.0. The overall trend might indicate a zero-point problem;
    a shift of all stars by 0.130\,dex gives a lower spread of the 
    calculated metallicities compared to adapted values, $\sigma_{+0.130}$\,=\,0.217.
    This is however still not satisfactory in comparison with the empirical equations which
    generally show smaller spread and imply higher metallicities (see Table \ref{tab:standDev} 
    and Figure \ref{fig:fitF}). A linear regression for the F-star calibration values to the adopted 
    values for the standard stars and the C06 sample is plotted in Figure \ref{fig:fitF}. The standard deviation 
    for this line fit is $\sigma$\,=\,0.172\,dex. 
    When applying our calibration to the C06 sample we see the same indications as for the 
    standard sample, except for the lower metallicities, where the theoretical calibration 
    suggest somewhat lower metallicities ($\sim$0.05\,dex) than derived in C06.
    However, the 13 F-stars in this sample are too few to
    allow any definite conclusions.
    When making use of the 266 F-stars in the VF sample and remembering that this sample 
    is highly biased towards metal-rich dwarf stars ([Fe/H]$>-0.2$ \& $\log g >4.0$) we 
    see the same trend as for the standard sample, i.e. our calibration suggests 
    lower metallicities than those given by VF05 (see also Table \ref{tab:standDev}).
        
    For the G-stars we find similar standard deviations for our standard sample with 
    respect to literature values, see Table \ref{tab:standDev}.
    There is an overall tendency that our
    theoretical calibration implies lower metallicities than listed in the
    literature for the metal-rich standard stars.
    The increasing deviation with increasing metallicity is of the same order
    as for the F-star calibration, i.e $\approx$\,0.3\,dex at [Me/H]\,=\,0.0.
    By adding 0.190\,dex to all stars we would obtain a lower spread, 
    $\sigma_{+0.190}$\,=\,0.181, which is of the same order as the deviation 
    shown by empirical equations (see Table \ref{tab:standDev}). 
    A linear regression to the calculated metallicities for the standard sample and C06 
    is plotted as a solid line in the figure ($\sigma$\,=\,0.200). 
    When using the 69 G-stars in the C06 sample, the result for the standard 
    sample is confirmed, the theoretical calibration suggests lower metallicities 
    for the higher metallicity range ([Fe/H] $> -0.5$) than listed in literature. 
    Applying the calibrations to the 694 G-stars in the VF05 sample, we obtain similar 
    results as for the C06 sample.

    For comparison the metallicities of our three comparison samples are calculated
    with 3 different empirical calibrations. The resulting standard deviations with
    respect to literature values can be seen in Table \ref{tab:standDev} and 
    linear regressions when applied to our standard sample can be seen in Figure \ref{fig:fitF} \&
    \ref{fig:fitG} for F-stars and G-stars, respectively.

    We note in passing that Clem et al. (2004) also found discrepancies between their
    calculated $m_{1}$ indices and observations. For their coolest models they had to
    apply upward corrections to their calculated $m_{1}$ indices of 0.1--0.3 mag. Our 
    $m_{1}$ indices depart considerably less from observations but some upward corrections
    would be needed to fit the coolest stars in Figure \ref{fig:m1vsbywithstars}.

    The strong effects of microturbulence on the strengths of the dominating lines in the $v$ band 
    makes it important to investigate whether this could be the reason for the mis-match of the 
    $m_1$ index in the theoretical calibration. In Table \ref{tab:microsens} we examine the 
    effects of changing the microturbulence parameter $\xi_{t}$. The changes have been chosen 
    to be larger for the lower-gravity models to take the larger parameter values usually 
    obtained by spectroscopy for giants into account. We see that the effect of increasing 
    the microturbulence is generally larger for lower $\log g$ and lower temperatures, as 
    expected. When comparing to Figure \ref{fig:m1vsbywithstars} we see that this effect 
    would indeed improve the situation by shifting the theoretical curves  
    and steepening the curves in the low $T_{\rm eff}$ end. Yet, the effects are considerably 
    smaller than those needed to eliminate the mis-match. 
 
    \begin{table}[h!]
      \caption{The change in the $uvby$ indices when the microturbulence parameter is increased} 
      \begin{tabular}{ccccc}
       \hline \hline
       $\log g$\,=\,4.0 & & $\Delta(\xi_{t}\,=\,1.0$ & $ \rightarrow$ & $\xi_{t}\,=\,1.5)$  \\
       $T_{\rm eff}$ &  [Me/H]   &   $\Delta c_1$     &   $\Delta m_1$ &    $\Delta (b-y)$   \\
       4500         &    0.00   &   0.0    &   0.04  &   0.011  \\
       5500         &    0.00   &   0.003  &   0.008 &   0.005 \\
       7000         &    0.00   &   0.004  &   0.005 &   0.001 \\
       4500         & $-$1.00   &   0.006  &   0.005 &   0.004 \\
       5500         & $-$1.00   &   0.006  &   0.005 &   0.002 \\
       7000         & $-$1.00   &   0.004  &   0.002 &   0.00  \\
       \hline
       $\log g$\,=\,3.0 & & $\Delta(\xi_{t}\,=\,1.0$ & $ \rightarrow$ & $\xi_{t}\,=\,2.0)$ \\ 
       $T_{\rm eff}$ &  [Me/H]   &  $\Delta c_1$      & $\Delta m_1$  &   $\Delta (b-y)$    \\
       4500         &     0.00  &  $-$0.016  & 0.023   &  0.032   \\ 
       5500         &     0.00  &     0.004  &  0.025  &  0.014   \\ 
       7000         &     0.00  &     0.001  &  0.014  &  0.002   \\ 
       4500         &  $-$1.00  &     0.015  &  0.018  &  0.012   \\ 
       5500         &  $-$1.00  &     0.017  &  0.013  &  0.003   \\ 
       7000         &  $-$1.00  &     0.006  &  0.005  &  0.0     \\ 
       \hline
      \end{tabular}
      \label{tab:microsens}
    \end{table} 
    
    Another circumstance which might have some significance as an explanation for the problems 
    with the $m_1$ sensitivity is the effect of lines from the (0,1) band of the CN violet 
    system in the $v$ band. In particular for the cooler giant stars, which may be affected 
    by the first dredge-up of CNO-processed material, the CN lines may become stronger due 
    to this; even if the carbon abundances are reduced by CNO processing, the enhanced N 
    abundance (to which C is converted) makes the CN lines stronger. Also the $^{13}$CN 
    lines should be significantly enhanced due to the production of $^{13}$C and N by CNO 
    processing. We have explored these effects by systematically increasing the N abundances 
    by a factor of 3, keeping the C abundance constant. This should lead to an overestimate 
    of the effect, except for possibly stars high-up on the giant branch. As indicated in 
    Figure \ref{fig:m1vsby} this only leads to some effects for the more metal-rich giant stars 
    and is not the explanation for the mis-match discussed here.   
    
   \section{Surface-gravity calibration}
   \label{sec:gravcalib}
\begin{figure*}[hbtp]
    \begin{center}
     \includegraphics[width=13cm,angle=90]{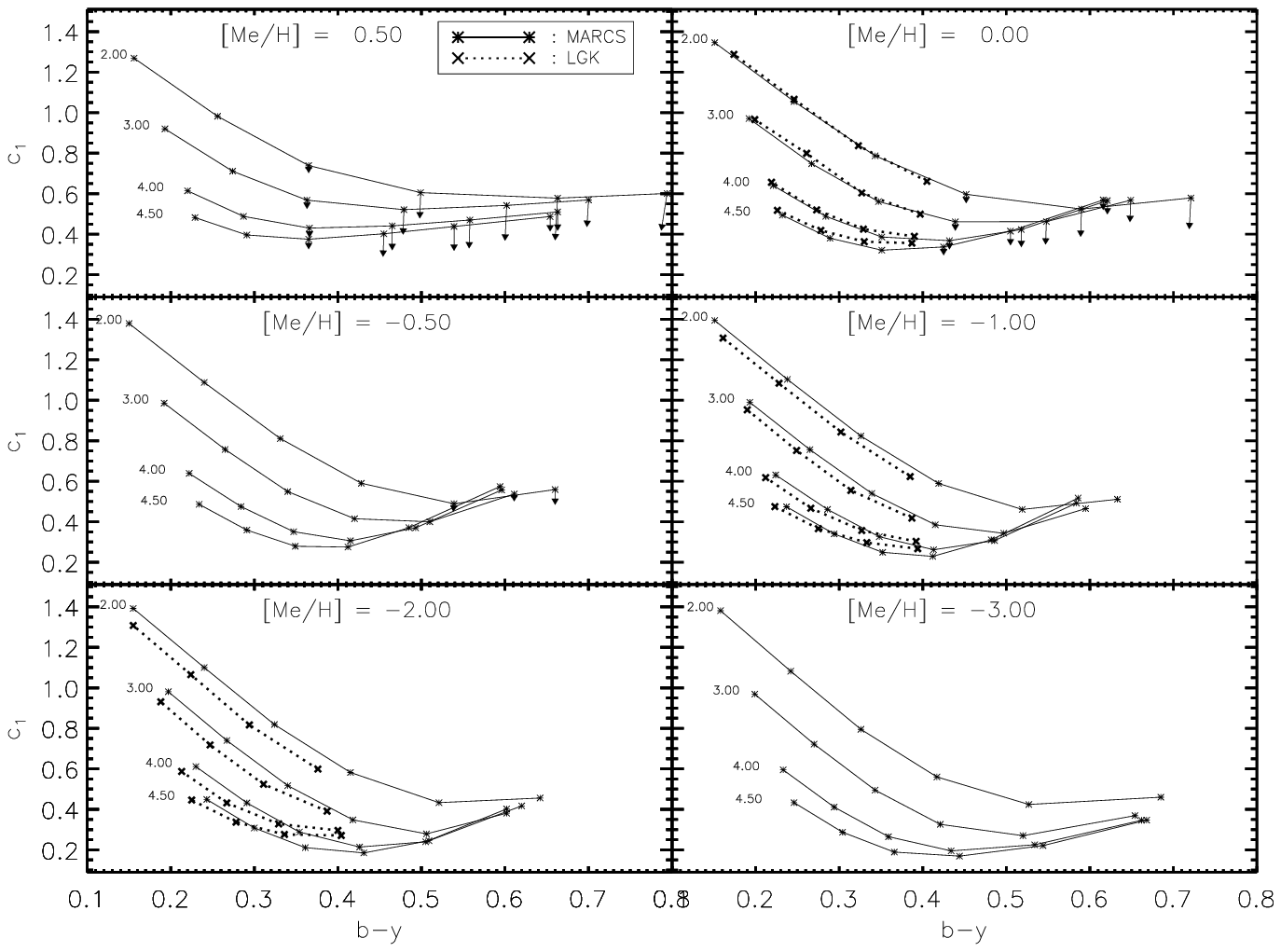}
     \caption{The $c_{1}$ index versus $(b-y)$ for MARCS and Lester, Gray \& Kurucz (1986, LGK)
              model atmospheres with different $\log g$ values indicated at corresponding curves 
              in the figure and different [Me/H] plotted in different panels. Arrows show the 
              effect of increasing the N abundances by a factor of 3.}
     \label{fig:c1vsby}
    \end{center}
   \end{figure*}

   The $c_1\equiv (u-v)-(v-y)$ index is designed to measure the Balmer discontinuity which 
   is a temperature indicator for B- and A-type stars and a surface-gravity indicator for the 
   late-type stars. Figure \ref{fig:c1vsby} shows its behaviour with changing parameters
   of the models. 
   Obviously, it works nicely as a gravity criterion, with some dependence
   on metallicity for the most metal-rich stars, which has been a disputed issue in earlier 
   calibration work. We note that for dwarfs cooler than the Sun it seems not very useful 
   as a gravity measure, while for bright giants, and not the least metal-poor ones, it 
   should work down to effective temperatures around 5000\,K. In Figure \ref{fig:c1vsby} we 
   have also plotted the indices calculated by Lester, Gray \& Kurucz (1986). In view of 
   the differences in line data and hydrogen-line theory we find the agreement satisfactory.

\begin{figure*}[hbtp]
    \begin{center}
     \includegraphics[width=13cm,angle=90]{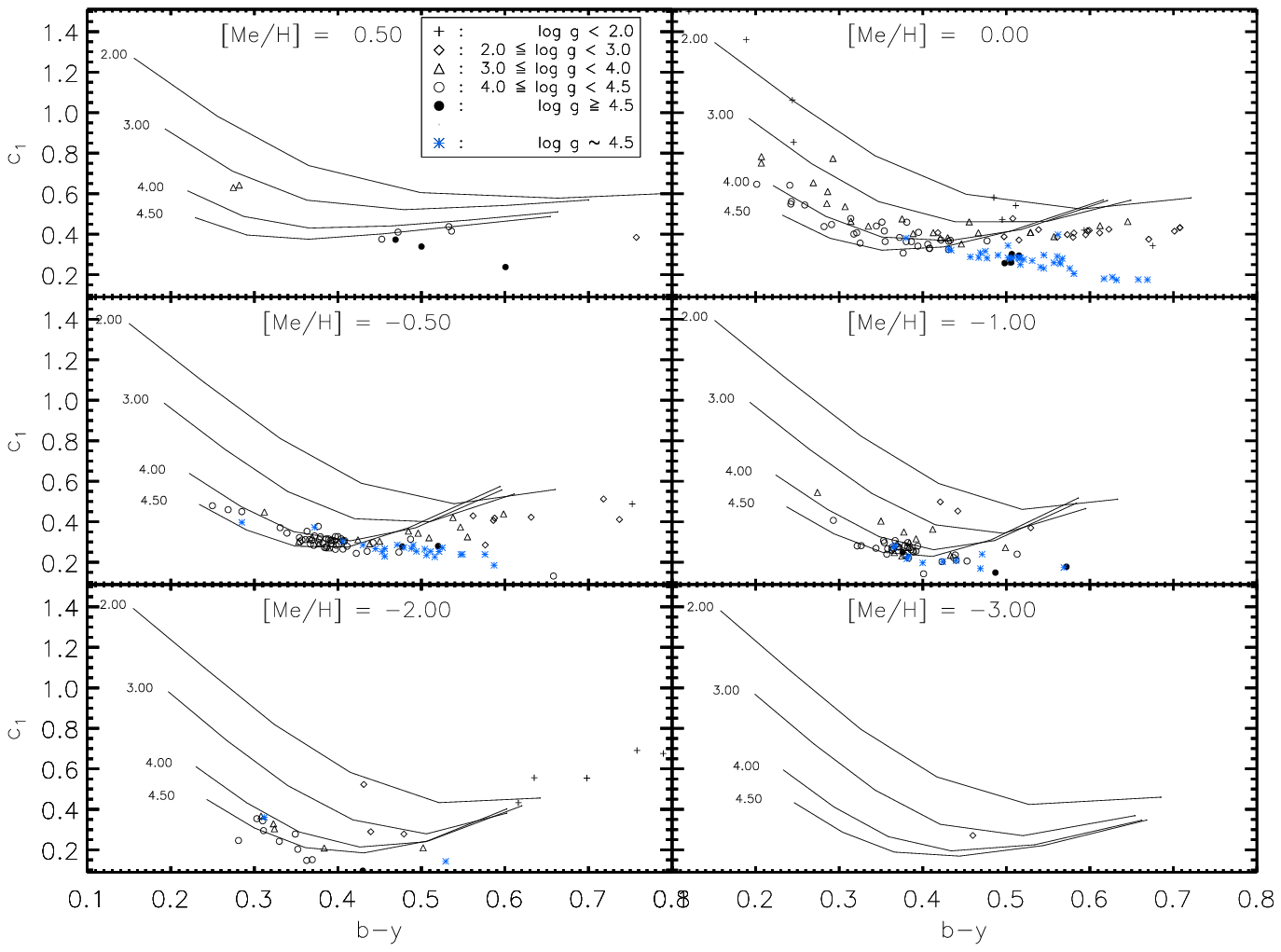}
     \caption{The $c_{1}$ index versus $(b-y)$ for MARCS models with different 
              [Me/H] and $\log g$ plotted together with values for 
              standard stars and Casagrande et al. (2006, C06) stars (marked as asterisks).
              The [Fe/H] intervals for the standard and C06 stars are [Fe/H]\,$\pm$0.1\,dex
              and [Fe/H]\,$\pm$0.2\,dex, respectively.}
     \label{fig:c1vsbywithstars}
   \end{center}
   \end{figure*}

   Analogously with our treatment of the metallicity dependence of the $m_{1}$
   index, we have measured the quantity $\delta c_1/\delta \log g$ at constant
   $(b-y)$ as a measure of the gravity sensitivity of $c_1$. Again we
   have to turn back to earlier calibrations to find corresponding direct
   empirical measures. Thus, Schuster \& Nissen (1989) have elaborated the methodology
   of Crawford (1975, 1979) and write
   \begin{eqnarray}
   \label{eqcalib2}
    M_V=M_{V,ZAMS}-f\cdot\delta c_0 \nonumber \\
    f\equiv 9.0+38.5\cdot((b-y)_0-0.22),
   \end{eqnarray}
   where $0.22\le (b-y)_0\le 0.47$, $(b-y)_0$ being the dereddened
   $(b-y)$, and $\delta c_0$ is the difference of a dereddened
   $c_1$ index and a standard sequence with $M_{V,ZAMS}$ at a given H$\beta$.
   From this one may estimate the empirical sensitivity
   $\delta c_1/\delta \log g$ to be approximately
   proportional to $2.5/f$. We thus obtain the following empirical values of 
   ($\delta c_1/\delta \log g;\,(b-y)$) for [Fe/H]\,=\,0: (0.21;0.30),
   (0.16;0.40) and (0.15;0.47). From Figure \ref{fig:c1vsby} and Table \ref{table:c1sensteff} we 
   measure the corresponding theoretical $\delta c_1/\delta \log g$
   values to be 0.19, 0.08 and 0.02, respectively.
   Thus, we find that the empirical gravity sensitivity of the $c_1$ index is well
   reproduced by the hotter models while it becomes underestimated for the cooler
   ones. We have also compared the calculated $c_1$ indices with the observed
   ones for our samples of standard stars and find a good agreement for the
   hotter stars (cf. Fig. \ref{fig:c1vsbywithstars}) while for the cooler stars there 
   is a severe mismatch.
   This is in itself not very remarkable -- the total line blocking in the
   ultraviolet spectra of the cooler stars is considerably greater than $70\%$,
   and it is to be expected that this will not be very accurately described by the
   model spectra. The fluxes in the $u$ and $v$ bands are highly sensitive to other
   parameters, such as microturbulence, and in particular CN abundance for the cooler giants, 
   and these are known to vary systematically with gravity. In fact, as is indicated in 
   Figure \ref{fig:c1vsby}, the CN line strengthening which is expected for the
   red giants improves the fit to observed $c_1$ indices for the red giant models considerably. 

   Clem et al. (2004) applied semiempirical corrections also to their calculated $(u-v)$ and 
   $(v-b)$ colours in order to fit observations. The resulting effect on their $c_{1}$ indices
   is typically less than 0.1 mag while we would need a downward correction of our $c_{1}$
   indices of about 0.2 mag for the coolest models to fit (cf. Fig. \ref{fig:c1vsbywithstars})
   the metal poor stars and even more for the metal-rich ones.

   \begin{table}[h!]
   \caption{The sensitivity to surface gravity, ($\delta c_1/\delta \log g)_{(b-y)}$, for 
            models with [Me/H]\,=\,$0.5$ and $0.0$} 
   \begin{tabular}{cccc}
     \hline \hline
     $(b-y)/\log g$ & 2.0--3.0 & 3.0--4.0 & 4.0--4.5 \\
     \hline
     $\rm [Me/H]$\,=\,0.5 & & &                       \\
     0.25 & 0.231 & 0.216 &  0.207 \\   
     0.30 & 0.213 & 0.187 &  0.165 \\  
     0.35 & 0.183 & 0.148 &  0.122 \\  
     0.40 & 0.148 & 0.112 &  0.091 \\  
     0.45 & 0.114 & 0.086 &  0.074 \\  
     0.50 & 0.083 & 0.071 &  0.060 \\  
     0.55 & 0.056 & 0.062 &  0.049 \\
          &        &        &                    \\
     $\rm [Me/H]$\,=\,0.0 & & &                  \\ 
      0.25 & 0.247 & 0.230 &   0.223 \\   
      0.30 & 0.236 & 0.205 &   0.185 \\   
      0.35 & 0.215 & 0.169 &   0.131 \\   
      0.40 & 0.182 & 0.126 &   0.082 \\   
      0.45 & 0.144 & 0.081 &   0.035 \\   
      0.50 & 0.106 & 0.040 & --0.006 \\   
      0.55 & 0.065 & 0.002 & --0.026 \\
     \hline
    \end{tabular}
    \label{table:c1sensteff}
   \end{table} 

   In a recent article, Twarog et al. (2007) discuss the metallicity dependence of the 
   $c_{1}$ index for disk stars ([Fe/H]\,$\geq$\,--1.00) and conclude that the
   metallicity sensitivity of the index generally has been underestimated. An interesting 
   test would therefore be to follow their recipe, i.e. divide our model grid into 
   three groups (hot, warm and cold), and check whether or not our synthetic colours 
   show the same behaviour. For the hot models ($(b-y)\,<\,0.43$) we find the same strong 
   metallicity dependence of the $m_{1}$ index and the same lack of sensitivity of the 
   $c_{1}$ index as is found by Twarog et al. For warm models (0.43\,$\leq\,(b-y)\,<$\,0.50) 
   we do find, like Twarog et al., a strong $c_{1}$ sensitivity to metallicity, but 
   likewise a relatively strong dependence of the $m_{1}$ index in contrast to the result 
   of Twarog et al. 
   For cool stars ($(b-y)\,\geq$\,0.50), we separate unevolved dwarfs from subgiants 
   and giants by making use of the defined LC ($\equiv c_{1} - 2.0m_{1} + 3.0(b-y) - 0.15)$ 
   calibration in Twarog et al. The models
   follow the same trends and LC separates the more metal-poor models ([Fe/H]\,=\,--0.5 \& --1.0) 
   but does not succed in separating the metal-rich ones ([Fe/H]\,$\geq\,0.0$), as is 
   also found by Twarog et al.
   It should however be noted that when applying the metallicity calibrations for hot and 
   warm stars as defined by Twarog et al. on our standard sample, we do not 
   reach any higher accuracy in reproducing the literature values of [Fe/H] than when 
   using e.g. S\&N89.

 \subsection{Applications}
  In the area of application of the $uvby$ photometry, some particular questions have been of special 
  interest to us: to which extent can the photometry be used for determining gravities and temperatures
  for metal-poor stars, and how sensitive is the synthetic $c_{1}$ index to certain elements
  such as nitrogen.

  Following Clem et al. (2004), we shall test our model colours versus observed colours for globular
  clusters. Unlike the approach of Clem et al., our main emphasis lies on exploring the possibilities
  and shortcomings of our theoretical model colours sooner than correcting them semiempirically to
  establish new calibrations.

  \subsubsection{M92}

  As a check on the capabilities of the $c_1$ index for determining gravities, we have tested the model 
  indices relative to the observed $uvby$ photometry for the extreme Pop II globular cluster M92.
  Adopting a metallicity of [Fe/H]=\,\,--2.22, a distance of 8.3 kpc, and a reddening of $E(B-V)=0.023$ 
  (Grundahl et al. 2000), 
  we have converted the observed $y-(b-y)$ diagram of Grundahl et al. (private communication)
  to the fundamental parameters along the evolutionary sequence past the turnoff point, 
  using $M_{v}$, $(b-y)$, $T_{\rm eff}$ (determined by MARCS models) and assuming a mass 
  of 0.8M$_{\odot}$ for the giants; c.f. the $T_{\rm eff}/\log g$ values given in 
  Table \ref{tab:M92}. The bolometric corrections used were taken from VandenBerg \& Clem (2003).   
  Next, we have calculated values for the $uvby$ colours using MARCS models for these parameter 
  values, to compare with the directly observed dereddened indices for M92 stars, according 
  to Grundahl et al. As is seen in Table \ref{tab:M92}, the observed colours 
  are well reproduced by the models. We have also calculated the derivatives 
  $\delta c_{1}/\delta \log g$ along the giant branch and find typical values ranging from 
  --0.2 to --0.1, thus agreeing within 20\,\% between the observations and the calculations. This 
  suggests that we can use the $uvby$ system to estimate gravities for metal-poor giant 
  stars.

\begin{table}[h!]
  \caption{Observed and calculated colours for $\log g$ values along the sub-giant - giant branch 
           for M92 and MARCS, respectively. The tabulated colours are mean colour values of
           stars close to the given fundamental parameters.} 
    \begin{tabular}{ccc|cc|cc}
       \hline \hline
              &      &        &  M92  &        & MARCS &    \\
       $\log g$ & $T_{\rm eff}$ &  $(b-y)_0$ & $c_0$ & $m_{0}$ & $c_{1}$ & $m_{1}$  \\
       \hline	    
       1.58   & 4682 &  0.584 & 0.516 &  0.109 & 0.519 &  0.117 \\
       2.05   & 4915 &  0.522 & 0.414 &  0.088 & 0.440 &  0.088 \\
       2.51   & 5125 &  0.473 & 0.374 &  0.075 & 0.384 &  0.067 \\
       2.99   & 5306 &  0.437 & 0.307 &  0.069 & 0.328 &  0.052 \\
       3.50   & 5525 &  0.400 & 0.281 &  0.057 & 0.295 &  0.047 \\
       3.72   & 5836 &  0.355 & 0.332 &  0.047 & 0.335 &  0.045 \\
       4.00   & 6302 &  0.299 & 0.383 &  0.064 & 0.411 &  0.055 \\
       \hline
    \end{tabular}
    \label{tab:M92}
\end{table}

  

 \subsubsection{NGC 6397}
  As a second test, effective temperatures for stars in the metal-poor globular 
  cluster NGC 6397 were derived. 
  This is an application of current interest as Korn et al. (2007) have
  recently claimed that they have identified abundance trends between groups of 
  cluster stars which they suggest are caused by atomic diffusion.
  Tracing abundance differences between different groups, at the turnoff,
  on the subgiant branch and on the lower and higher red-giant branch
  (TOP, SGB, lRGB, hRGB), requires well determined stellar parameters,
  and most importantly well measured effective
  temperature differences between the groups. Korn et al. derive effective
  temperatures with a photometric and spectroscopic approach and 
  they find consistent temperatures. 
  One potential problem, however, is the fact that two different photometric
  calibrations needed to be applied (Alonso et al. 1996, 1999) which may 
  introduce systematic errors.
  For our theoretical calibration we derive
  effective temperatures for each of the 18 stars in the four 
  different groups, by using $\log g$ values and dereddened colour indices 
  $(b-y)$ and $(v-y)$ listed in Korn et al. and adopting their [Fe/H] of --2.0.
  Mean values of the effective temperatures for the four groups are shown in 
  Table \ref{tab:NGC6397}.
  The results indicate that our effective temperatures are 50\,K to 100\,K hotter than 
  obtained from the empirical calibrations. More importantly, the temperature differences 
  between TOP, SGB and hRGB stars are close to (even somewhat smaller than) the ones
  obtained from the empirical calibrations by Korn et al. This strengthens the 
  claim of Korn et al. that abundance differences in Li, Mg and Fe, reflecting 
  atomic diffusion, are present at a significant level.

  \begin{table}
    \caption{Derived effective temperatures from empirical calibrations
      of photometry ($(b-y)$ and $(v-y)$), and spectroscopy (Korn et al. 2007), as well as 
      our theoretical calibration, for groups of stars in the metal-poor cluster NGC 6397. The 
      temperatures are mean values for a number of stars/models in each group.}
    \begin{tabular}{c|ccccc}
      \hline \hline
                        & Phot.   &         & Spec. & MARCS  &                \\
      $T_{\rm eff}$\,[K] & $(b-y)$ & $(v-y)$ &       &$(b-y)$ & $(v-y)$        \\
      \hline	                          
      hRGB              & 5121    &  5132   & 5130  & 5225  &  5226  \\
      lRGB              & 5455    &  5408   & 5456  & 5541  &  5535  \\
      SGB               & 5797    &  5824   & 5805  & 5839  &  5853  \\
      TOP               & 6229    &  6214   & 6254  & 6288  &  6281  \\
      \hline                                          
      $\Delta$(TOP-RBG) & 1108    &  1081   & 1124  & 1063  &  1055  \\
      $\Delta$(TOP-SGB) & 433     &  390    & 449   & 449   &  428   \\

    \end{tabular}
    \label{tab:NGC6397}
  \end{table}

  \subsubsection{NGC 6752}
  In another test,
  the nitrogen sensitivity of of the $c_{1}$ index was
  examined. In a recent paper, Yong et al. (2008), measure the nitrogen content of
  giants in the globular cluster NGC 6752 ([Fe/H]\,$\sim$\,--1.6) based on
  high-resolution observations of NH. The authors find a strong correlation between
  nitrogen content and the $c_{1}$ index and establish the $c_{y}$ index ($c_{1}-(b-y)$)
  that shows a close to linear correlation with respect to nitrogen content.
  To examine the synthetic sensitivity of this $c_{y}$ index, a set of
  models and high resolution spectra were calculated ($T_{\rm eff}$ = 4749, 4829, 
  4841, 4904, 4950\,K, $\log g$ = 1.95, 2.10, 2.15, 2.33, 2.42, [Me/H] = --1.50) based
  on a selection of stars discussed by Yong et al. The [N/Fe] content was set to vary as 
  0.00, 0.75 and 1.50 for each of the models in order to cover a great part of the 
  [N/Fe] span in NGC 6752 found by Yong et al. As a result we find that our synthetic
  $c_{y}$ index is sensitive to the nitrogen content of the model, even though the 
  sensitivity $\delta c_{y}/\delta$[N/Fe] of the models is somewhat smaller, 
  $\sim0.03-0.04$ (see Table \ref{tab:yongtest}), 
  compared to the sensitivity Yong et al. find for their giant stars, $\sim$0.06, 
  as is estimated from the best-fit straight line in Yong et al. Figure 7.

  \begin{table}
    \caption{Calculated $\delta c_{y} / \delta$[N/Fe] for giant models as a
             comparison to the observed $c_{\rm y}$ sensitivity of NGC 6752 
             (Yong et al. (2008))}
    \begin{tabular}{cc|cc}
      \hline \hline
       $T_{\rm eff}$ & $\log g$  & 0.0$--$0.75   &  0.75$--$1.5 \\
      \hline	                          
       4749         & 1.95      &    0.03      &    0.00     \\
       4829         & 2.10      &    0.04      &    0.02     \\
       4841         & 2.15      &    0.04      &    0.02     \\
       4904         & 2.33      &    0.04      &    0.04     \\
       4950         & 2.42      &    0.04      &    0.04     \\ 
    \end{tabular}
    \label{tab:yongtest}
  \end{table}

\section{Conclusions and recommendations}

We have explored the possibilities and shortcomings of synthetic $uvby-$H$\beta$ photometry
based on new MARCS model atmospheres. In general a good agreement with
empirical calibrations of this photometric system is found. However, a number of systematic
deviations between theory and observations also become apparent. The temperature
sensitivity of the $(b-y)$ colour (i.e. $\delta (b-y)/\delta T_{\rm eff}$) 
seems marginally larger for the calculated colours than
is found when using infrared-flux-method determinations of temperatures.
A similar, and even somewhat larger, difference occurs for temperatures based 
on the H$\beta$ index, when compared with the empirical scale.
The $(b-y)$ calibration for Pop II stars supports Korn et al. (2007), who claim the signatures
of atomic diffusion in the metal-poor globular cluster NGC 6397.

For the metallicity index $m_1$ the theoretical sensitivity $\delta m_1/\delta $[Me/H] is 
somewhat larger than the empirical one that is based on spectroscopic [Fe/H]-determinations. 
For the gravity sensitivity of the $c_1$ Balmer discontinuity measure, we find a reasonably 
good agreement with observations for stars hotter than the Sun, where the Balmer discontinuity
is significant. Considerable problems remain for the cooler stars, although the model
calibration works well for Pop II giants.

One might ask whether these problems may be solved when even more detailed atomic and molecular
line data become available to feed into model atmospheres. This is possible but it is also possible
that thermal inhomogenities as generated by convection and possibly also non-LTE effects, 
contribute to the 
difference between synthetic and observed values of $(b-y)$ and $m_1$. Concerning the problems 
in reproducing the observed $c_1$ indices of cooler stars, these may well be due to failures 
in line-data, although thermal inhomogeneities in the atmospheres may also here turn out to 
be the important effect. One should also realize that there are still residual problems
associated with differences and uncertainties in the observed and adopted Str\"omgren passbands
(Manfroid \& Sterken, 1987).
The current study illustrates the problems in synthetic photometry in the visual wavelength regions.
Continued efforts along such lines must be complemented with more detailed 
studies of the shortcoming of classical model atmospheres and the replacement of those, also
in large-scale calibration efforts, by physically more realistic models.     

\acknowledgements{Anders Eriksson is thanked for a major contribution 
  at the startup of this project. Remo Collet is thanked for discussions on
3D model atmospheres and Ulrike Heiter and Andreas Korn for valuable suggestions and 
comments on the manuscript. Ana Garc\'{i}a P\'{e}rez is thanked for the calculations of reddening.}

\thebibliography{}{
\bibliographystyle{astron}
\bibliography{mnemonic,ref_C}
\bibitem[]{}
Adelman, S.J., Gulliver, A.F. 1990, ApJ 348, 712
\bibitem[]{} 
Ali, A. W., \& Griem, H. R. 1966, Phys. Rev. 144, 366
\bibitem[]{}
Alonso, A., Arribas, S., Martinez-Roger, C. 1996, A\&A 313, 873
\bibitem[]{}
Alonso, A., Arribas, S., Martinez-Roger, C. 1999, A\&AS 140, 261
\bibitem[]{}
Ardeberg, A., Lindgren H. 1981, RMxA\&A 6, 173
\bibitem[]{}
Asplund, M., Grevess, N., Sauval, A.J. 2005, ASPC 336, 25
\bibitem[]{}
Barklem, P. S., \& Piskunov, N. 2003, in Modelling of Stellar
Atmospheres, ed. N. Piskunov, W. W. Weiss, D. F. Gray, Proc., IAU
Symp. 210
\bibitem[]{}
Baschek, B. 1960, Zeitschrift Ap 50, 296
\bibitem[]{}
Bell, R.A. 1970, MNRAS 148, 25
\bibitem[]{} 
Bell, R.A. 1971, MNRAS 154, 343
\bibitem[]{}
Bell, R.A., Parsons, S.B. 1974, MNSAS 169, 71
\bibitem[]{}
Bessell, M.S. 2005, A\&A Rev., 43, 293  
\bibitem[]{}
Casagrande, L., Portinari, L., Flynn, C. 2006, MNRAS 373, 13
\bibitem[]{}
Castelli, F., Kurucz, R.L. 2006, A\&A 454, 333
\bibitem[]{}
Christlieb, N., Gustafsson, B., Korn, A.J. et al. 2004, ApJ 603, 708
\bibitem[]{}
Clem, J.L., VandenBerg, D.A., Grundahl, F., Bell, R., 2004, AJ 127, 1227
\bibitem[]{}
Crawford, D.L. 1958, ApJ 128, 185 
\bibitem[]{}
Crawford, D.L. 1966, IAU Symp. 24, 170
\bibitem[]{} 
Crawford, D.L. 1975, AJ 80, 955
\bibitem[]{}
Crawford, D.L. 1979, AJ 84, 1858
\bibitem[]{}
Crawford, D.L., Mander, J. 1966, AJ 71, 114 
\bibitem[]{}
Crawford, D.L., Barnes, J.V. 1970, AJ 75 978
\bibitem[]{}
Edvardsson, B. 2008, in "A stellar journey. A symposium in celebration of Bengt Gustafsson's
65th birthday", Physica Scripta, in press
\bibitem[]{}
Edvardsson, B., Eriksson, K., Gustafsson, B. et al. 2008, A\&A, in preparation
\bibitem[]{}
Gigas, D. 1988, A\&A 192, 264
\bibitem[]{}
Grevesse, N., Sauval, A.J. 1998, SSRv 85, 161
\bibitem[]{}
Griem, H. R. 1960, ApJ 132, 883
\bibitem[]{}
Grundahl, F., VandenBerg, D.A., Bell, R.A., Andersen, M.I., Stetson, P.B. 2000, 
A.J. 120, 1884 
\bibitem[]{}
Gulliver, A.F., Hill, G., Adelman, S.J. 1994, ApJ 429, L81
\bibitem[]{}
Gustafsson, B., Bell, R.A. 1979, A\&A 74, 313
\bibitem[]{}
Gustafsson, B., Edvardsson B., Eriksson K., J\o rgensen U.G., Plez B. 2003,
Conf. Ser. Vol. 288, p.331
\bibitem[]{}
Gustafsson, B., Edvardsson B., Eriksson K., et al. 2008, A\&A in press, arXiv:0805.0554v1
\bibitem[]{}
Gustafsson, B., Nissen, P.E. 1972, A\&A 19, 291
\bibitem[]{}
Hauck, B. \& Mermilliod, M. 1998, A\&AS 129, 431
\bibitem[]{}
Hakkila, J., Myers, J.M., Stidham, B.J., \& Hartmann, D.H. 1997, AJ 114, 2043
\bibitem[]{}
Heiter, U., Weiss, W.W., Paunzen, E. 2002, A\&A 381, 971
\bibitem[]{}
Helt, B.E., Franco G.A.P., Florentin Nielsen R., 1987, in: ESO Workshop on the SN1987A, 
Danziger I.J.(ed.), Garching, p. 89 
\bibitem[]{}
Hill, G.M. 1995, A\&A 294, 536
\bibitem[]{}
Hill, G., Gulliver, A.F., Adelman, S.J. 2004, IAU Symposium 224, 35
\bibitem[]{}
Hoffleit, D., Warren, W.H. Jr. 1995, VizieR On-line Data Catalog: V/50, SIMBAD
\bibitem[]{}
Holmberg, J., Nordstr\"om, B., Andersen, J. 2007, A\&A 475, 519
\bibitem[]{}
Ilijic, S., Rosandic, M., Dominis, E., et al. 1998, CoSka 27, 467
\bibitem[]{}
Jonsell, K., Edvardsson, B., Gustafsson, B. et al. 2005, A\&A 440, 321
\bibitem[]{}
Korn, A.J., Grundahl, F., Richard, O., Mashonkina, L., Barklem, P.S., Collet, R.,
Gustafsson, B., Piskunov, N. 2007, ApJ 671, 402
\bibitem[]{}
Kupka, F., Piskunov, N. E., Ryabchikova, T.A., et al. 1999, A\&AS, 138,119
\bibitem[]{}
Kurucz, R.L. 1995, Kurucz CD-ROM No.15, Cambridge, Mass., Smithsonian Astrophysical Observatory
\bibitem[]{}
Lester, J.B., Gray, R.O., Kurucz, R.L. 1986, ApJS 61, 509
\bibitem[]{}
Lejeune, T., Lastennet, E., Westera, P., Buser, R. 1999, ASP Conf Proc. 192, 207
\bibitem[]{}
Manfoid, J. \& Sterken, C. 1987, A\&AS 71, 539
\bibitem[]{}
Nissen, P.E. 1970, A\&A
\bibitem[]{}
Nissen, P.E. 1981, A\&A 97, 145
\bibitem[]{}
Nissen, P.E. 1988, A\&A 199, 146
\bibitem[]{}
Nissen, P.E., Gustafsson, B. 1978, in Astronomical Papers Dedicated to Bengt Str\"omgren, 
  Copenhagen 1978, p. 43 
\bibitem[]{}
Nordstr\"om, B., Mayor, M., Andersen, J. et al. 2004, A\&A 418, 989
\bibitem[]{}
Olsen, E.H. 1983, A\&AS 54, 55
\bibitem[]{}
Olsen, E.H. 1984, A\&A 57, 443
\bibitem[]{}
Olsen, E.H. 1988, A\&A 189, 173
\bibitem[]{}
Paunzen, E., Iliev, I.K., Kamp, I. et al. 2002, MNRAS 336, 1030
\bibitem[]{}
Plez, B., Masseron, T., Van Eck, S. et al. 2008, ASP, Conf. Ser., in presss
\bibitem[]{}
Querci, F., Querci, M., Kunde, V. 1971, A\&A 31, 265
\bibitem[]{}
Ram\'{i}rez, I., Mel\'{e}ndez, J. 2005, ApJ 626, 465
\bibitem[]{}
Ram\'{i}rez, I., Mel\'{e}ndez, J. 2005, ApJ 626, 446
\bibitem[]{}
Relyea, L.J., Kurucz, R.L. 1978, ApJS 37, 45
\bibitem[]{}
Schuster, W.J., Nissen, P.E. 1988, A\&AS 73, 225
\bibitem[]{}
Schuster, W.J., Nissen, P.E. 1989, A\&A 221, 65
\bibitem[]{}
Str\"omgren, B. 1963, QJRAS 4, 8
\bibitem[]{}
Str\"omgren, B. 1964, Astrophys. Norveg. 9, 333
\bibitem[]{}
Str\"omgren, B. 1966, Ann Rev Astr Ap 4, 433
\bibitem[]{}
Twarog, B.A., Vargas, L.C., Anthony-Twarog, B.J. 2007, AJ 134, 1777
\bibitem[]{}
VandenBerg, D.A., Clem, J.L. 2003, AJ 126, 778
\bibitem[]{}
Valenti, J.A., Fischer, D.A. 2005, ApJS 159, 141
\bibitem[]{}
Vidal, C. R., Cooper, J., \& Smith, E. W. 1973, ApJS 25, 37
\bibitem[]{}
Wallerstein, G. 1962, ApJS 6, 407
\bibitem[]{}
Yong, D., Grundahl, F., Johnson, J.A., Asplund, M. 2008, ApJ 684, 1159
}

\newpage

\begin{appendix}
  \section{Calibrations}
  In the following $T_{\rm eff}$ and [Fe/H] calibrations we have systematically 
  used the form, and in most cases the index limits for the validity of the 
  calibrations, according to Alonso et al. (1996,1999) and Schuster \& Nissen (1989),
  respectively. The coefficients have then been calculated by a least square method to 
  optimize the fit to the model indices.

\subsection{Theoretical $T_{\rm eff}$ calibration for dwarf stars, $(b-y)$ and H$\beta$}
\label{sec:app_teffByBetaCalib}
$\Theta_{\rm eff} \equiv 5040/T_{\rm eff}$

\begin{eqnarray*}
  \Theta_{\rm eff}(b-y)_{\rm dwarf} & = &  0.466 + 1.025(b-y) - 0.123(b-y)^{2} \\
         & &  + 0.212(b-y)c_{1} - 0.051(b-y)\rm [Me/H]        \\ 
         & &  + 0.003\rm [Me/H] - 0.005\rm [Me/H]^{2}
\end{eqnarray*}

The models, upon which the calibration is based, are selected
to follow the restrictions given for the empirical equation
(A96, Eq. 9. Restrictions adopted from Fig. 11a): 
$-3.00 \leq$ [Me/H] $\leq 0.50$; 
$0.70 \leq (b-y) \leq 0.20$; 
$0.10 \leq c_{1} \leq 0.55$; 
$4.00 \leq  \log g$ (set as limit for dwarf stars).\\\\

\noindent
\begin{eqnarray*}
 \Theta_{\rm eff}(\rm H\beta)_{\rm dwarf} & = & 
       28.60 - 19.79\rm H\beta + 3.504\rm H\beta^{2} + 0.422\rm H\beta\rm [Me/H] \\
       & &  - 1.068\rm [Me/H] + 0.002\rm [Me/H]^{2}
\end{eqnarray*}

In analogy with the theoretical calibration based on $(b-y)$, the
limits determining which models to use, were set to be equal to the
ones for empirical calibration (A96, Eq. 10 and
its following applicable ranges):
$2.44 \leq  $\,H\,$\beta  \leq 2.74$  for $ +0.5 \geq $\,[Me/H]\,$ > -0.5$; 
$2.50 \leq  $\,H\,$\beta  \leq 2.70$  for $ -0.5 \geq $\,[Me/H]\,$ > -1.5$; 
$2.50 \leq  $\,H\,$\beta  \leq 2.63$  for $ -1.5 \geq $\,[Me/H]\,$ > -2.5$; 
$2.51 \leq  $\,H\,$\beta  \leq 2.62$  for $ -2.5 \geq $\,[Me/H]\,$ > -3.5$; 
$4.0 \leq\ \log g$ (set as a limit for dwarf stars). 


\subsection{Theoretical $T_{\rm eff}$ calibration for giants,  $(b-y)$}
\label{sec:app_teffByCalibGiant}
\begin{eqnarray*}
 \Theta_{\rm eff}(b-y)_{\rm giant} & = & 
             a_{0} + a{_1}(b-y) + a{_2}(b-y){^2} + \\
             & & a{_3}(b-y)\rm [Me/H] + a{_4}\rm [Me/H] + a{_5}\rm [Me/H]{^2}
\end{eqnarray*}

\begin{table}[h!]
  \caption{The derived coefficients for the theoretical calibration in the intervalls
    I: $0.15 \leq (b-y) \leq 0.424$,  II: $0.424 \leq (b-y) \leq 0.712$ \&
       $-5.0 \leq  $ [Me/H] $ \leq -0.5$ and III: $0.428 \leq (b-y) \leq 0.794$ \&
       $-0.5 <  $ [Me/H] $ \leq 0.5$}
 \begin{center}
 \begin{tiny}
 \begin{tabular}{ccccccc}
  \hline \hline
       &  a$_{0}$        &  a$_{1}$ &  a$_{2}$ &  a$_{3}$  &  a$_{4}$  &  a$_{5}$      \\
         \hline                                                                      
   I:  &  0.6732         &   0.0859 & 1.1455  & -1.080e-2 & -0.132e-2 & -0.082e-2    \\
       &                 &          &         &           &           &              \\
   II: &  0.1983         & 2.0931   & -0.9978 & 4.709e-2  & -2.66e-2  & -0.01e-2    \\
       &                 &          &         &           &           &              \\
  III: &  0.4522         & 1.1745   & -0.3093 & -0.1693   &  2.165e-2 & -1.679e-2    \\

   \hline
  \end{tabular}
 \end{tiny}
 \end{center}
 \label{byfitparam}
\end{table}

Here we deviate from the same restriction limits for the calibrations as presented 
in Alonso et al (1999, Table 2 and 3). New theoretical limits were set due to
model restrictions: $(b-y)\,\leq\,0.424$ and $(b-y)\,\geq\,0.424$ (cf. Alonso et al 1999,
Eq. 14: $(b-y)\,\leq\,0.55$ and Eq. 15: ($b-y)\,\geq\,0.50$). The $\log g$ limit was
set to $1.5 \le \log g \le 3.5$.  

\subsection{Theoretical metallicity  calibration for G-stars} 
\label{sec:app_Gstar}
The models, upon which the metallicity calibrations for G and F stars are based, 
are selected to follow the restrictions given for the empirical equations
of S\&N89, Eq. 3. and 2., respectively. The G-stars limits cover the following
intervals:
$-2.6 \leq$ [Me/H] $\leq 0.4$; 
$0.37 \leq (b-y) \leq 0.59$; 
$0.03 \leq c_{1} \leq 0.57$;
$3.00 \leq  \log g$ (set as limit for $\log g$).\\
\begin{eqnarray*}
  \rm [Me/H]_{G} & = & -2.796 + 39.21m_{1} - 88.97m^{2}_{1} - 73.43m_{1}(b-y) \\
                 &   &  + 181.4m^{2}_{1}(b-y) + [27.03m_{1} - 1.220c_{1} - 41.42m^{2}_{1}]c_{1}  
\end{eqnarray*}

\subsection{Theoretical metallicity calibration for F-stars} 
\label{sec:app_Fstar}
F-star limits are given by: 
$-3.5 \leq$ [Me/H] $\leq 0.2$; 
$0.22 \leq (b-y) \leq 0.38$; 
$0.17 \leq c_{1} \leq 0.58$;
$3.00 \leq  \log g$ (set as limit for $\log g$).\\
\begin{eqnarray*}
  \rm [Me/H]_{F} & = & 1.850 - 34.21m_{1} + 105.34m_{1}(b-y)                  \\
                 &   & + 179.8m^{2}_{1}(b-y) - 242.4m_{1}(b-y)^{2}             \\
                 &   & + [2.757 - 20.38m_{1} + 0.2777(b-y)]\log(m_{1} - c_{3}) \\
                 &   &                                                        \\
  c_{3}          & = & 0.4462 - 2.233(b-y) + 2.885(b-y)^{2}    
\end{eqnarray*}

%

\end{appendix}

\end{document}